\documentclass[twocolumn]{pasj01}
  \usepackage{color}
  
\begin{document}
  \Received{2019/04/14}%{yyyy/mm/dd}
  \Accepted{2019/12/19}%{yyyy/mm/dd} 
  %\Published{yyyy/mm/dd}
  
  \title{Evidence for planetary hypothesis for PTFO\,8-8695\,b with five-year optical/infrared monitoring observations}
  
  %%% begin:list of authors
  % Do NOT capitalize all letters in "textsc".
  \author{Yuta \textsc{Tanimoto}\altaffilmark{1, 2,}$^{*}$%
  %\thanks{Example: Present Address is xxxxxxxxxx}
  }
  \email{yuta.tanimoto@nao.ac.jp}
  
  \author{Takuya \textsc{Yamashita}\altaffilmark{2}}
  
  \author{Takahiro \textsc{Ui}\altaffilmark{3}}

  \author{Mizuho \textsc{Uchiyama}\altaffilmark{2}}
  
  \author{Miho \textsc{Kawabata}\altaffilmark{4, 3}}
  \author{Hiroki \textsc{Mori}\altaffilmark{3}}
  \author{Tatsuya \textsc{Nakaoka}\altaffilmark{5, 3}}
  
  \author{Taisei \textsc{Abe}\altaffilmark{3}}
  \author{Ryosuke \textsc{Itoh}\altaffilmark{6, 3}}
  \author{Yuka \textsc{Kanda}\altaffilmark{3}}
  \author{Kenji \textsc{Kawaguchi}\altaffilmark{3}}
  \author{Naoki \textsc{Kawahara}\altaffilmark{3}}
  \author{Ikki \textsc{Otsubo}\altaffilmark{3}}
  \author{Kensei \textsc{Shiki}\altaffilmark{3}}
  \author{Kengo \textsc{Takagi}\altaffilmark{3}}
  \author{Katsutoshi \textsc{Takaki}\altaffilmark{3}}

  \author{Hiroshi \textsc{Akitaya}\altaffilmark{5}}
  \author{Masayuki \textsc{Yamanaka}\altaffilmark{4, 5, 7}}
  \author{Koji S. \textsc{Kawabata}\altaffilmark{5}}
  
  \altaffiltext{1}{Department of Astronomy, The University of Tokyo, 7-3-1 Hongo, Bunkyo-ku, Tokyo 113-0033, Japan}
  \altaffiltext{2}{National Astronomical Observatory of Japan, 2-21-1 Osawa, Mitaka, Tokyo 181-8588, Japan}
  \altaffiltext{3}{Graduate School of Science, Hiroshima University, 1-3-1 Kagamiyama, Higashi-Hiroshima, Hiroshima 739-8526, Japan}
  \altaffiltext{4}{Department of Astronomy, Kyoto University, Kitashirakawa-Oiwake-cho, Sakyo-ku, Kyoto 606-8502, Japan}
  \altaffiltext{5}{Hiroshima Astrophysical Science Center, Hiroshima University, 1-3-1 Kagamiyama, Higashi-Hiroshima, Hiroshima 739-8526, Japan}
  \altaffiltext{6}{Bisei Astronomical Observatory, 1723-70 Okura, Bisei, Ibara, Okayama 714-1411, Japan}
  \altaffiltext{7}{Okayama Observatory, Kyoto University, 3037-5 Honjo, Kamogata, Asakuchi, Okayama 719-0232, Japan}
  %%% end:list of authors
  
  %% `\KeyWords{}' always has to be placed before `\maketitle'.
   %Do NOT move this preamble from here!
  
  \maketitle 
  
  \begin{abstract}
    PTFO\,8-8695\,b (CVSO\,30\,b) is a young planet candidate whose host star is a $\sim 2.6\>$Myr-old T-Tauri star, and there have been continuous discussions about the nature of this system.
    To unveil the mystery of this system, we observed PTFO\,8-8695 for around five years at optical and infrared bands simultaneously using Kanata telescope at the Higashi-Hiroshima Observatory.
    Through our observations, we found that the reported fading event split into two: deeper but phase-shifted ``dip-A'' and shallower but equiphase ``dip-B''.
    These dips disappeared at different epochs, and then, dip-B reappeared.
    Based on the observed wavelength dependence of dip depths, a dust clump and a precessing planet are likely origins of dip-A and B, respectively.
    Here we propose ``a precessing planet associated with a dust cloud'' scenario for this system.
    This scenario is consistent with the reported change in the depth of fading events, and even with the reported results, which were thought to be negative evidence to the planetary hypothesis, such as the past non-detection of the Rossiter--McLaughlin effect.
    If this scenario is correct, this is the third case of a young ($<3\>$Myr) planet around a pre-main sequence star.
    This finding implies that a planet can be formed within a few Myr.
  \end{abstract}
  
  \section{Introduction}
    Confirming over 4000 exoplanets, the existence of a significant number of planets quite unlike the solar system, such as ``hot Jupiters,'' has been revealed.
    This wide variety has been raising our interest in the way of forming planets.
    To explain the formation of gas giants, the following two models have been mainly proposed: the core accretion model (e.g., \cite{Hayashi1985, Pollack1996}) and the gravitational instability model (e.g., \cite{Boss1997}).
    The former argues that an up-to-$10\,M_\oplus$ core can be formed by an accumulation of planetesimals and then it accretes surrounding gas to become a gas giant.
    This model is thought to be the standard scenario for the formation process of the solar system and is supported by the planet-metallicity correlation (e.g., \cite{Gonzalez1997}).
    However, this suffers from the closeness of timescales of the core formation and gas depletion, which may result in an inability to attach enough gas to form a gas envelope around the core.
    On the other hand, the gravitational instability model claims that planets are formed in a disk fragment through self-gravitational instability of the disk.
    The result of simulation conducted by \citet{Boss1997}, according to which planets can be formed in $\sim 10^3\>\mathrm{yr}$ at the outer region of the disk, around $8\>\mathrm{au}$ is consistent with the existence of giant planets far from their host stars detected by direct imaging (e.g., HR~8799~b, c, d, and e orbits farther than $11\>\mathrm{au}$: \cite{Marois2008}, \yearcite{Marois2010}).
    
    Although both these models have some supporting evidence as noted, neither can produce hot Jupiters by itself.
    Therefore, an orbital evolution mechanism, such as disk-planet interaction (e.g., \cite{Lin1996}), planet-planet scattering \citep{Chatterjee2008a}, and Kozai--Lindov mechanism \citep{Kozai1962}, is necessary, which makes the whole process more complicated.
    To limit the timescale of planet formation would be the key to solve this difficulty.
    In other words, detecting and confirming young hot Jupiters potentially make a significant contribution to reveal the planet formation process.
 
    PTFO\,8-8695\,b was reported as the first hot Jupiter candidate around a pre-main-sequence star in \citet{VanEyken2012}.
    Its host star, PTFO\,8-8695 also called as CVSO\,30, is a weak-lined T-Tauri star (WTTS) located in the Orion-OB1a region, and its age is estimated to be around $2.6\>$Myr \citep{Briceno2005}.
    \citet{VanEyken2012} observed this object for two years from 2009 to 2010 as part of the Palomar Transient Factory Orion project \citep{VanEyken2011} and successfully extracted periodic transit signals from the stellar variability often observed in T-Tauri stars.
    However, the main problem was that the shapes of the detected transits were changing and asymmetric, unlike the ``ordinary'' planetary transits.

    A combination of gravity darkening and orbital precession was proposed by \citet{Barnes2013} to explain these odd features.
    Gravity darkening was predicted by \citet{Zeipel1924} at first.
    A rapidly rotating star yields brighter regions around poles and a fainter region around the equator, because of its oblate shape.
    They claimed that the observed asymmetry can be explained by this unevenness of brightness at the stellar surface.
    Furthermore, changing transit paths through nodal precession results in long-term variability of the transit shape.
    This model was tuned up later by \citet{Kamiaka2015} to agree with their new observations.

    Then, \citet{Ciardi2015} found that transits changed in depth, disappeared, and reappeared, and that the Rossiter--McLaughlin (RM) effect was not detected from their new follow-up observations.
    Although they claimed that the temporary disappearance was supporting evidence to the precession model, their data were not completely consistent with the gravity-darkening model proposed by \citet{Barnes2013}.
    This disagreement threw doubt on the planetary hypothesis.
    \citet{Yu2015} found that the planetary hypothesis was unfavorable to explain their three main results: continuous detections of the fading event without cessation through their monitoring inconsistent with the precession model, absence of secondary eclipse in both Spitzer $4.5\>\micron$ data and Magellan \textit{H-}band data, and non-detection of the RM effect.
    In addition, \citet{Howarth2016} disfavored this hypothesis because physically implausible values were necessary for a combination of the precession and their modified gravity darkening to reproduce even light curves observed by \citet{VanEyken2012}.
    
    Although once the planetary hypothesis had no edge, \citet{Raetz2016} and \citet{Johns-Krull2016} supported the presence of a disintegrating planet.
    The former monitored for three years and could not confirm the shrinking period reported in \citet{Yu2015}.
    Because \citet{Yu2015} denied the hypothesis of a disintegrating planet based on this fast orbital decay, their observations suggested that this hypothesis was still possible.
    The latter observed $\mathrm{H\alpha}$-line profiles and detected an occasionally appearing excess component, which periodically shifts in wavelength.
    Because the radial velocities of this excess $\mathrm{H\alpha}$ were consistent with the velocity positions of the planetary companion predicted from \citet{VanEyken2012}, they concluded that this phenomenon was caused by the mass outflow from the planet. They also noted that the mass outflow had been suppressed because of host star's flares when there was no excess component.

    The latest study of this system was conducted by \citet{Onitsuka2017}.
    One transit in 2016 was observed at three optical bands simultaneously, which had shallower depths at longer wavelengths.
    They claimed that a transiting dust clump or occultation of an accretion hotspot was a remaining possibility because of the wavelength dependence of the depth.

    An argument about the nature of the phenomena has been lasting as above, and no conclusion is still settled.
    Thus, continuous observations are important to reveal the nature of this mysterious object.
    Here, we report the results of five-year monitoring of fading events at optical and infrared bands simultaneously to investigate the nature of this mysterious object.
    Our observations are shown in section 2 and we describe the photometry method and following light curve analysis, such as transit fitting and updating ephemeris, in section 3. 
    Section 4 presents an interpretation of the origins of the fading events.
       
  \section{Observations}
    We observed PTFO\,8-8695 for 20 days from 2014 February 23 to 2018 December 30 with Hiroshima Optical and a near-infrared camera (HONIR: \cite{Akitaya2014}) and Hiroshima one-shot wide-field polarimeter (HOWPol: \cite{Kawabata2008}) on a 150-cm Kanata telescope at the Higashi-Hiroshima Observatory.

    HONIR is located on the Cassegrain focus and can take optical and near-infrared images simultaneously.
    Although it has capabilities of spectroscopic and polarimetric observations, only an imaging mode was employed for our observations.
    Its optical arm is equipped with a fully depleted CCD array with $2048 \times 4096\>\mathrm{pixels}$ but only $2048 \times 2048\>\mathrm{pixels}$ are used in order to match its field of view and pixel scale with those of the infrared arm which is equipped with a HgCdTe detector with $2048 \times 2048\>\mathrm{pixels}$.
    A modified Johnson--Cousins filter system (i.e., \textit{B, V}, $R_C$, and $I_C$ bands) is mounted on the optical arm, and the Mauna Kea Observatory’s filter system (i.e., \textit{J, H} and $K_s$ band) is mounted on the infrared arm.
    Its field of view is about $10' \times 10'$ and the pixel scale is about $0.3\>\mathrm{arcsec\>pixel^{-1}}$ at both optical and infrared arms.

    HOWPol is a wide-field optical polarimeter located on the Nasmyth focus.
    Observations were conducted only in the imaging mode as HONIR observations, and its field of view is about $15'$ diameter and the pixel scale is about $0.3\>\mathrm{arcsec\>pixel^{-1}}$ in this mode.
    HOWPol is composed of two fully depleted CCDs with $2048 \times 4096\>\mathrm{pixels}$, and the modified Johnson-Cousins filter system is mounted on HOWPol as the optical arm of HONIR.

    Every observation was conducted without dithering,
    that is, fixing stellar images at the same positions on the detectors in order to minimize the systematic error originating from imperfect flat correction.
    We used only the $I_C$ band for the optical observations and \textit{J, H} and $K_s$ bands for the infrared observations.
    The details of each observation are listed in table~\ref{tab:obs}.
    \begin{table*}
      \tbl{List of observations}{
      \begin{tabular}{lllccp{0.01pt}ccc} \hline
        &&& \multicolumn{2}{c}{Optical} && \multicolumn{2}{c}{Infrared} & Number of exposures \\ \cline{4-5} \cline{7-8}
        UT Date & UT Time & Instrument & Filter & Exposure time$\>$(s)\footnotemark[$*$] && Filter & Exposure time$\>$(s)\footnotemark[$*$] & for each filter \\  \hline
        2014 Feb 23 & 10:48--14:33 & HONIR  & $I_C$ & 110 && $K_s$   & 95 & 103 \\
        2014 Dec 27 & 11:53--17:19 & HONIR  & $I_C$ & 110 && $K_s$   & 95 & 147 \\
        2015 Jan 10 & 09:32--15:08 & HONIR  & $I_C$ & 110 && $K_s$   & 95 & 152 \\
        2015 Jan 23 & 11:11--15:25 & HONIR  & $I_C$ & 110 && \textit{J} & 95 & 115 \\
        2015 Feb 10 & 09:27--14:11 & HONIR  & $I_C$ & 75  && \textit{H} & 60 & 173 \\
        2015 Feb 14 & 09:46--15:10 & HONIR  & $I_C$ & 110 && $K_s$   & 95 & 142 \\
        2015 Feb 23 & 10:08--13:32 & HONIR  & $I_C$ & 110 && $K_s$   & 95 & 93  \\
        2015 Oct 17 & 15:22--20:21 & HOWPol & $I_C$ & 30  &&--&--& 420 \\
        2016 Nov 3  & 15:27--19:54 & HONIR  & $I_C$ & 110 && $K_s$   & 95 & 118 \\
        2016 Nov 25 & 14:42--19:47 & HONIR  & $I_C$ & 110 && \textit{J} & 95 & 130 \\
        2016 Nov 29 & 11:27--20:10 & HONIR  & $I_C$ & 110 && \textit{J} & 95 & 201 \\
        2016 Dec 1  & 12:12--17:32 & HONIR  & $I_C$ & 110 && \textit{J} & 95 & 124 \\
        2016 Dec 9  & 12:26--18:01 & HONIR  & $I_C$ & 110 && \textit{J} & 95 & 142 \\
        2017 Oct 3  & 16:37--20:10 & HONIR  & $I_C$ & 110 && \textit{J} & 95 & 81  \\
        2018 Feb 8  & 12:12--15:55 & HONIR  & $I_C$ & 110 && \textit{J} & 95 & 96  \\
        2018 Nov 7  & 15:04--20:48 & HOWPol & $I_C$ & 30  &&--&--& 380 \\
        2018 Nov 9  & 13:14--20:52 & HOWPol & $I_C$ & 30  &&--&--& 641 \\
        2018 Nov 10 & 14:05--20:28 & HOWPol & $I_C$ & 30  &&--&--& 536 \\
        2018 Dec 29 & 12:08--15:55 & HONIR  & $I_C$ & 110 && \textit{J} & 95 & 97 \\
        2018 Dec 30 & 10:36--14:19 & HONIR  & $I_C$ & 110, 95\footnotemark[$*$] && \textit{J} & 95, 80\footnotemark[$\dagger$] & 115 \\
        \hline
      \end{tabular}}\label{tab:obs}
      \begin{tabnote}
        \footnotemark[$*$] Exposure times of the optical band is longer than infrared bands. This is because readout of the infrared image completes within a shutter aperture time whereas that of the optical image is equal to the aperture time. \\
        \footnotemark[$\dagger$] Exposure time was changed in the middle of the observations.
      \end{tabnote}
    \end{table*}  
    Bias frames (for optical), dark frames (for Infrared) and dome flat frames (for both) were taken on the same day as (or adjacent days to) the observation.
    Sky images, for sky subtraction, were taken in the same field as the target but with dithering before and after the target observations.  
    Raw images were reduced using HONIR and HOWPol pipeline which include trimming, bias/dark subtraction, flat fielding, and sky subtraction. 
  
  \section{Data analysis and results}
    \subsection{Photometry}
\begin{figure}
        \begin{center}
          \includegraphics[width=8cm]{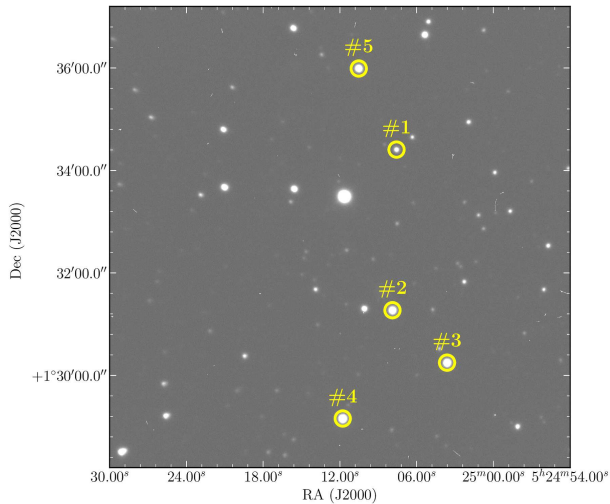} 
        \end{center}
        \caption{$I_C$-band image obtained by HONIR. \#1 is PTFO\,8-8695. \#2 and \#3 are reference stars for optical images, while the four stars from \#2 to \#5 are those for infrared images.}\label{fig:refstar}
      \end{figure}

      To investigate the variations in stellar brightness, relative photometry was performed.
      We derived the relative flux of a star from $f_\mathrm{obj}/(\prod_{i=1}^n \sqrt[n]{f_i})$, where $f_\mathrm{obj}$ and $f_i$ are ADU values of the object and an $i$-th reference star estimated from the circular aperture photometry of \verb|SExtractor| \citep{Bertin1996}.
      The \textit{S/N} ratio of the reference stars do not depend much on their brightness they are bright enough and their count variation mainly depends on their variability.
      So, we treated them equally by setting the denominator as not an arithmetical average but a geometrical average.
      To get the best accuracy for the relative flux, we should choose appropriate photometric parameters: a selection of reference stars and an aperture size.
      Good comparison stars are non-variable but many of the stars in our field can be variable because the observing field, 25 Ori, is a very young cluster.
      So, we derived rms variations of the "light curve" of ADU count ratios for every combination of all detected stars in the field and chose several stars as our comparison stars based on the rms variations.  
      The adopted comparison stars are shown in figure~\ref{fig:refstar}.
      We adopted aperture sizes for each day by calculating apparent \textit{S/N} ratios while changing an aperture size.
      The adopted aperture radius for each day range over $11$--$20\>\mathrm{pixel}$ for HONIR optical images, $10$--$16\>\mathrm{pixel}$ for HONIR infrared ones, and $7$--$10\>\mathrm{pixels}$ for HOWPol optical ones.
        
\begin{figure*}
        \begin{center}
          \includegraphics[width=16cm]{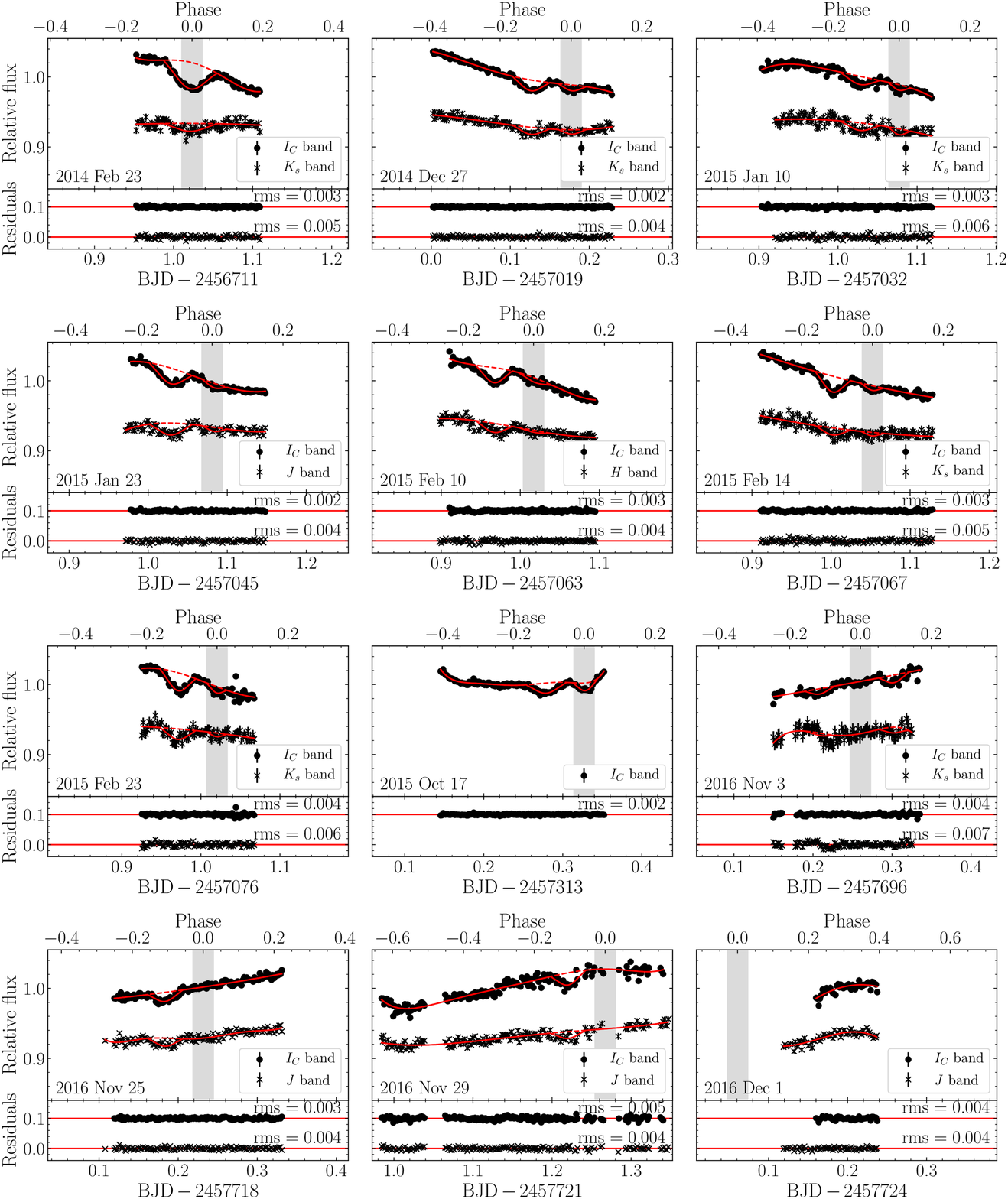}        
        \end{center}
        \caption{Obtained light curves of PTFO\,8-8695 from February 2014 to December 2016. The solid lines indicate the best-fitting light curve models and dashed lines indicate the trend functions. Gray-shaded areas indicate the expected time windows of dip-B (see subsection~\ref{sec:ephemeris}), whose centers are calculated from equation~(\ref{eq:ephemeris}) and widths are fixed to a minimum value, $0.026\>\mathrm{d}$ of the duration in table~\ref{tab:allparams}. The lower panels of each light curve show residuals from the best-fitting models.}\label{fig:rawlc_1}
      \end{figure*}

\begin{figure*}
        \begin{center}
          \includegraphics[width=16cm]{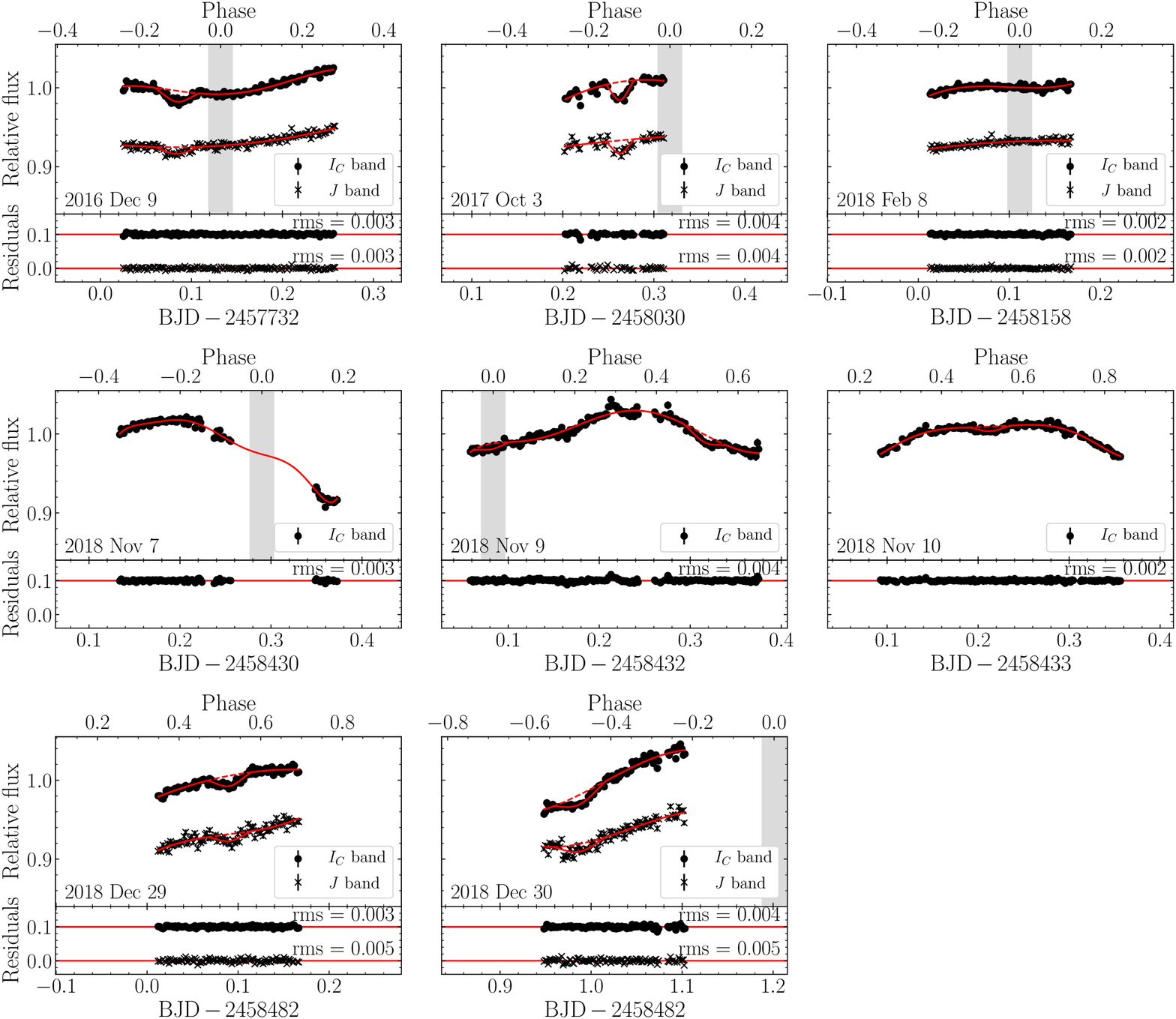}        
        \end{center}
        \caption{Same as figure~\ref{fig:rawlc_1}, but for light curves from December 2016 to December 2018.}\label{fig:rawlc_2}
      \end{figure*}

      Finally, transforming the time stamps from Julian Date to $\mathrm{BJD_{TDB}}$ using \verb|astropy.time| package, a \verb|Python|-based algorithm for dealing with times \citep{TheAstropyCollaboration2013,TheAstropyCollaboration2018}, we obtained the light curves shown in  figures~\ref{fig:rawlc_1} and~\ref{fig:rawlc_2}.
      The typical precision of the light curves is 0.2\% for the optical and 0.4\% for the infrared data.
      Note that HOWPol data were binned with $\sim 130\>\mathrm{s}$ width because their exposure time, $30\>\mathrm{s}$, is shorter than HONIR's exposure times, $75$--$110\>\mathrm{s}$ (see also table~\ref{tab:obs}).
  
    \subsection{Light curve fitting} \label{sec:fitting}
      The light curves obtained in figures~\ref{fig:rawlc_1} and~\ref{fig:rawlc_2} show two variability components, a stellar one and a transit-like dip, as noted in \citet{VanEyken2012}.
      Although out-of-transit light curves were fitted to extract dips in almost all previous studies of this object (e.g., \cite{VanEyken2012, Ciardi2015}), this method is sensitive to an artificially defined time window \citep{Rodenbeck2018}.
      Moreover, it was very difficult to set a correct time window because of the strong stellar variability trend and a possible split of a fading event, which we detected for the first time (see later discussion).
      Thus, we fitted the light curves with both trends and dips simultaneously.

      We assumed that the observed light curves are represented by the product of a polynomial trend and a dip function, $F_\mathrm{dip}$, which is defined by
      \begin{equation}\label{eq:f_tr}
        F_\mathrm{dip}=\left\{
        \begin{array}{ll}
          1 + \delta - 2\delta\mathrm{sech}\left(\frac{2.634(t-t_c)}{w}\right) & \mbox{if in transit,} \\
          1 & \mbox{otherwise,}
        \end{array}
        \right.
      \end{equation}
      where $\delta, w, t_c$ are the depth, duration from ingress to egress, and time of the center of a dip.
      Although a physically constructed transit model may be a better choice for $F_\mathrm{dip}$, it is still in a discussion whether a transiting planet causes this phenomenon.
      Furthermore, this step aims to distinguish dips from stellar variabilities and to monitor long-term behavior of dips (i.e., depth, center, and width).
      Given these two reasons, we employed the above dip function.

      The simultaneous fitting was performed as follows:
      \begin{enumerate}
        \item 
        Prepare 18 light curve models from combinations of six trend functions from 2nd-order to 7th-order polynomials and three types of dip functions, which have zero, one, and two dips.

        \item 
        Fit the optical light curves with all prepared models based on the Bayesian information criterion (BIC: \cite{Schwarz1978}). 
        BIC is defined as $\mathrm{BIC} \equiv \chi^2 + k\ln\,N$, where $k$ and $N$ are the numbers of free parameters and data points, respectively.
        The reason why we did not fit the infrared light curves together with the optical ones is that the infrared data had lower \textit{S/N} ratios than the optical ones.
        With this fitting, we extracted dip periods, which were used as fixed values in the following.

        \item
        By fitting the optical and infrared data with the masked dip periods, derive all best-fitting trend polynomials that show the lowest BIC among the polynomials from 2nd- to 7th-order.

        \item
        Flatten the light curves by dividing them by the derived best-fitting trend polynomial and derive the dip depths by fitting the flattened light curves with $F_\mathrm{dip}$.
      \end{enumerate}
      The best-fitting parameters of the dip function are listed in table~\ref{tab:allparams} and the best-fitting light curves are shown as solid lines in figures~\ref{fig:rawlc_1} and~\ref{fig:rawlc_2}.
      Note that although the data obtained on 2016 November 3 shows another dip at around phase 0.1, we do not claim detection of this possible dip here because it is detected only once.
      Moreover, apparently the infrared data on 2016 November 3 suffers from an extra noise and the fitting may not be correctly done (see figure~\ref{fig:rawlc_1}).
      So we exclude the data from following discussions.
      However, of course, the dip at around phase 0.1 can be real.
      It may be one of the emerging and distinguishing multiple dips of this object (see later discussions).

\begin{figure*}
        \begin{center}
          \includegraphics[width=16cm]{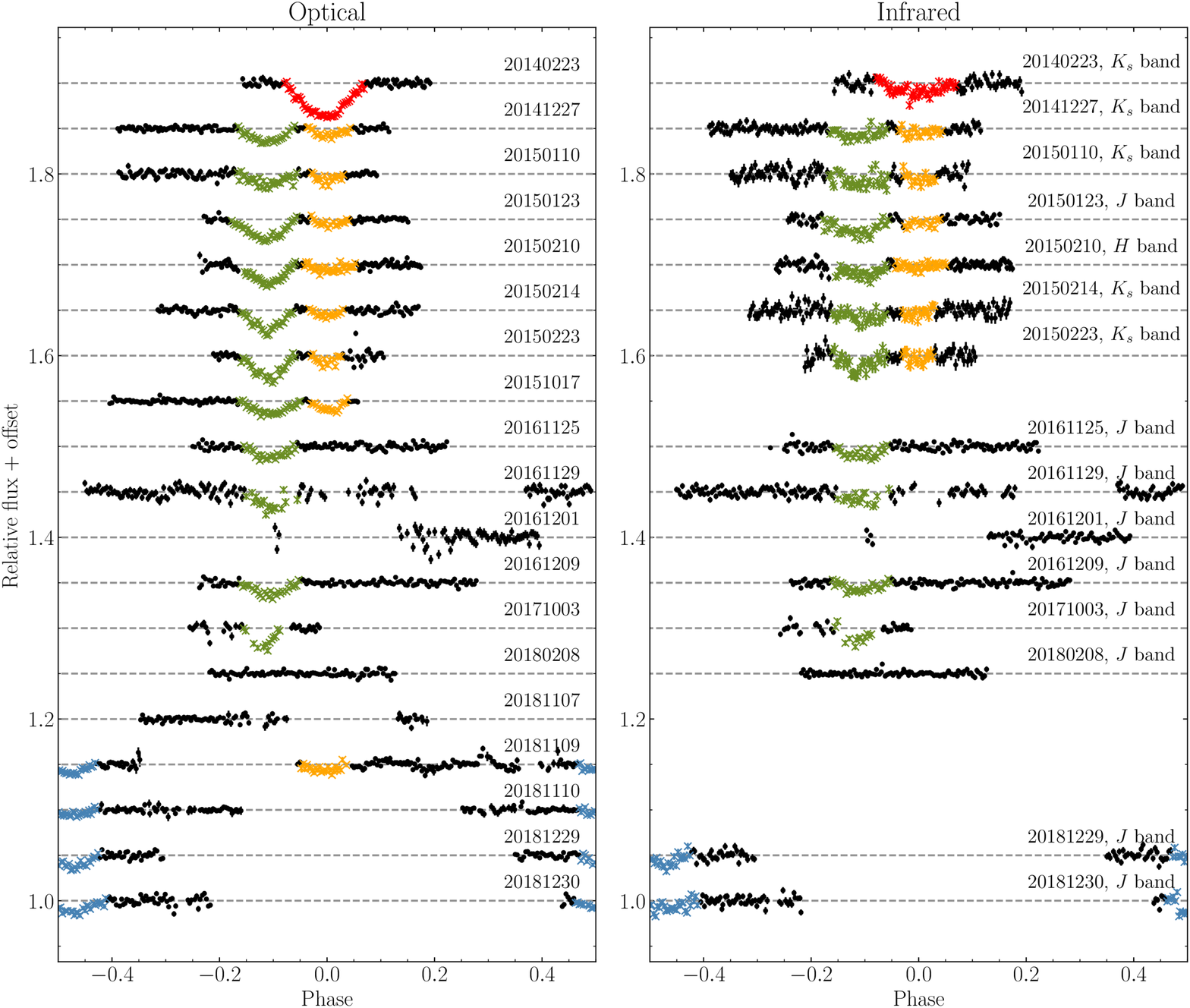}
        \end{center}
        \caption{All flattened light curves. The left panel shows optical light curves, while the right panel shows infrared ones. Circles are out-of-transit data while crosses are in-transit data. Red, green, orange, and blue crosses are "fading events before splitting", "dip-A", dip-B", and "dip-C", respectively (see discussions in the text). UT dates of each observation are shown at the upper right of each baseline. Phase is calculated using our new ephemeris, equation~(\ref{eq:ephemeris}).}\label{fig:lc_def}
      \end{figure*}

      All flattened light curves are shown in figure~\ref{fig:lc_def}.
      The models with two dips were adopted for some light curves as the best-fitting ones.
      From the aspect of time variation, one dip observed in February 2014 seems to have split into two since December 2014.
      The earlier dip (hereafter ``dip-A''), located on $\sim -0.1$ in phase (see figure~\ref{fig:lc_def}), was deeper than the later dips (hereafter ``dip-B'') but slightly shallower than the single one observed on 2014 February 23. 
      While dip-A lasted until October 2017, dip-B disappeared in October 2015 and reappeared from  November 2018.
      Moreover, the other dips at $\sim -0.45$ in phase (hereafter ``dip-C'') appeared on 2018 November 9.
      Note that these phases are based on equation~(\ref{eq:ephemeris}) in the next section.
      \begin{table*}
      \tbl{Best-fitting parameters of the dip function.}{
      \begin{tabular}{lcccp{0.01pt}cc}
        \hline & \multicolumn{3}{c}{From optical data} & & From infrared data & \\ \cline{2-4} \cline{6-6}
        UT Date & Center & Duration & Depth & & Depth & Dip type \\
        & $t_c\>(\mathrm{BJD}-2456000)$ & $w\>\mathrm{(d)}$ & $\delta\>(\%)$ & & $\delta\>(\%)$ & \\ \hline
        2014 Feb 23&  $712.0212\,\pm\,0.0005$& $0.068\,\pm\,0.002$& $3.73\,\pm\,0.18$& & $1.13\,\pm\,0.19$ & before splitting\\  \hline
        2014 Dec 27& $1019.1261\,\pm\,0.0007$& $0.047\,\pm\,0.003$& $1.54\,\pm\,0.09$& & $1.00\,\pm\,0.14$ & dip-A\\ 
        ibid.& $1019.1782\,\pm\,0.0010$& $0.036\,\pm\,0.004$& $0.85\,\pm\,0.10$& & $0.62\,\pm\,0.16$ & dip-B\\  \hline
        2015 Jan 10& $1033.0278\,\pm\,0.0012$& $0.051\,\pm\,0.005$& $1.18\,\pm\,0.12$& & $1.09\,\pm\,0.17$ & dip-A\\ 
        ibid.& $1033.0772\,\pm\,0.0013$& $0.029\,\pm\,0.005$& $0.81\,\pm\,0.15$& & $0.85\,\pm\,0.20$ & dip-B\\  \hline
        2015 Jan 23& $1046.0285\,\pm\,0.0005$& $0.055\,\pm\,0.002$& $2.20\,\pm\,0.12$& & $1.70\,\pm\,0.14$ & dip-A\\ 
        ibid.& $1046.0837\,\pm\,0.0014$& $0.032\,\pm\,0.005$& $0.63\,\pm\,0.11$& & $0.58\,\pm\,0.13$& dip-B\\  \hline
        2015 Feb 10& $1063.9668\,\pm\,0.0005$& $0.047\,\pm\,0.002$& $2.12\,\pm\,0.10$& & $1.28\,\pm\,0.12$& dip-A\\ 
        ibid.& $1064.0187\,\pm\,0.0015$& $0.043\,\pm\,0.006$& $0.62\,\pm\,0.09$& & $0.27\,\pm\,0.11$& dip-B\\  \hline
        2015 Feb 14& $1068.0021\,\pm\,0.0005$& $0.045\,\pm\,0.002$& $2.26\,\pm\,0.09$& & $1.02\,\pm\,0.20$& dip-A\\ 
        ibid.& $1068.0519\,\pm\,0.0013$& $0.029\,\pm\,0.005$& $0.66\,\pm\,0.11$& & $0.50\,\pm\,0.22$& dip-B\\  \hline
        2015 Feb 23& $1076.9709\,\pm\,0.0005$& $0.046\,\pm\,0.002$& $2.52\,\pm\,0.14$& & $1.49\,\pm\,0.25$& dip-A\\ 
        ibid.& $1077.0197\,\pm\,0.0010$& $0.026\,\pm\,0.004$& $0.89\,\pm\,0.14$& & $0.54\,\pm\,0.28$& dip-B\\  \hline
        2015 Oct 17& $1313.2793\,\pm\,0.0008$& $0.052\,\pm\,0.003$& $1.45\,\pm\,0.11$& & Not observed & dip-A\\ 
        ibid.& $1313.3283\,\pm\,0.0009$& $0.032\,\pm\,0.003$& $1.10\,\pm\,0.16$& & Not observed & dip-B\\  \hline
        2016 Nov 3 & $1696.2083\,\pm\,0.0021$& $0.038\,\pm\,0.008$& $0.76\,\pm\,0.17$& & $0.33\,\pm\,0.38$ & dip-A\\ 
        ibid. & $1696.3023\,\pm\,0.0016$& $0.039\,\pm\,0.006$& $1.09\,\pm\,0.17$& & $0.80\,\pm\,0.45$ & another dip\\  \hline
        2016 Nov 25& $1718.1835\,\pm\,0.0007$& $0.045\,\pm\,0.003$& $1.30\,\pm\,0.08$& & $1.15\,\pm\,0.13$ & dip-A\\ \hline 
        2016 Nov 29& $1722.2204\,\pm\,0.0013$& $0.045\,\pm\,0.005$& $1.71\,\pm\,0.15$& & $0.96\,\pm\,0.14$ & dip-A\\ \hline 
        2016 Dec 1 & \multicolumn{6}{c}{Not detected} \\ \hline 
        2016 Dec 9 & $1732.0837\,\pm\,0.0007$& $0.049\,\pm\,0.003$& $1.39\,\pm\,0.09$& & $0.85\,\pm\,0.08$ & dip-A\\ \hline 
        2017 Oct 3 & $2030.2628\,\pm\,0.0006$& $0.031\,\pm\,0.002$& $2.30\,\pm\,0.19$& & $1.66\,\pm\,0.17$ & dip-A\\ \hline 
        2018 Feb 8 & \multicolumn{6}{c}{Not detected} \\ \hline 
        2018 Nov 7 & \multicolumn{6}{c}{Not detected} \\ \hline
        2018 Nov 9 & $2432.0809\,\pm\,0.0018$& $0.041\,\pm\,0.007$& $0.73\,\pm\,0.14$& & Not observed & dip-B\\ 
        ibid. & $2432.3154\,\pm\,0.0011$& $0.047\,\pm\,0.004$& $0.98\,\pm\,0.10$& & Not observed & dip-C\\ \hline 
        2018 Nov 10& $2433.2130\,\pm\,0.0017$& $0.046\,\pm\,0.007$& $0.58\,\pm\,0.10$& & Not observed & dip-C\\ \hline 
        2018 Dec 29& $2482.0905\,\pm\,0.0007$& $0.048\,\pm\,0.003$& $1.32\,\pm\,0.09$& & $0.88\,\pm\,0.14$ & dip-C\\ \hline 
        2018 Dec 30& $2482.9874\,\pm\,0.0011$& $0.058\,\pm\,0.004$& $1.24\,\pm\,0.14$& & $0.97\,\pm\,0.16$ & dip-C\\ \hline
      \end{tabular}}\label{tab:allparams}
      \end{table*}

    \subsection{Ephemeris} \label{sec:ephemeris}
\begin{figure}
        \begin{center}
        \includegraphics[width=8cm]{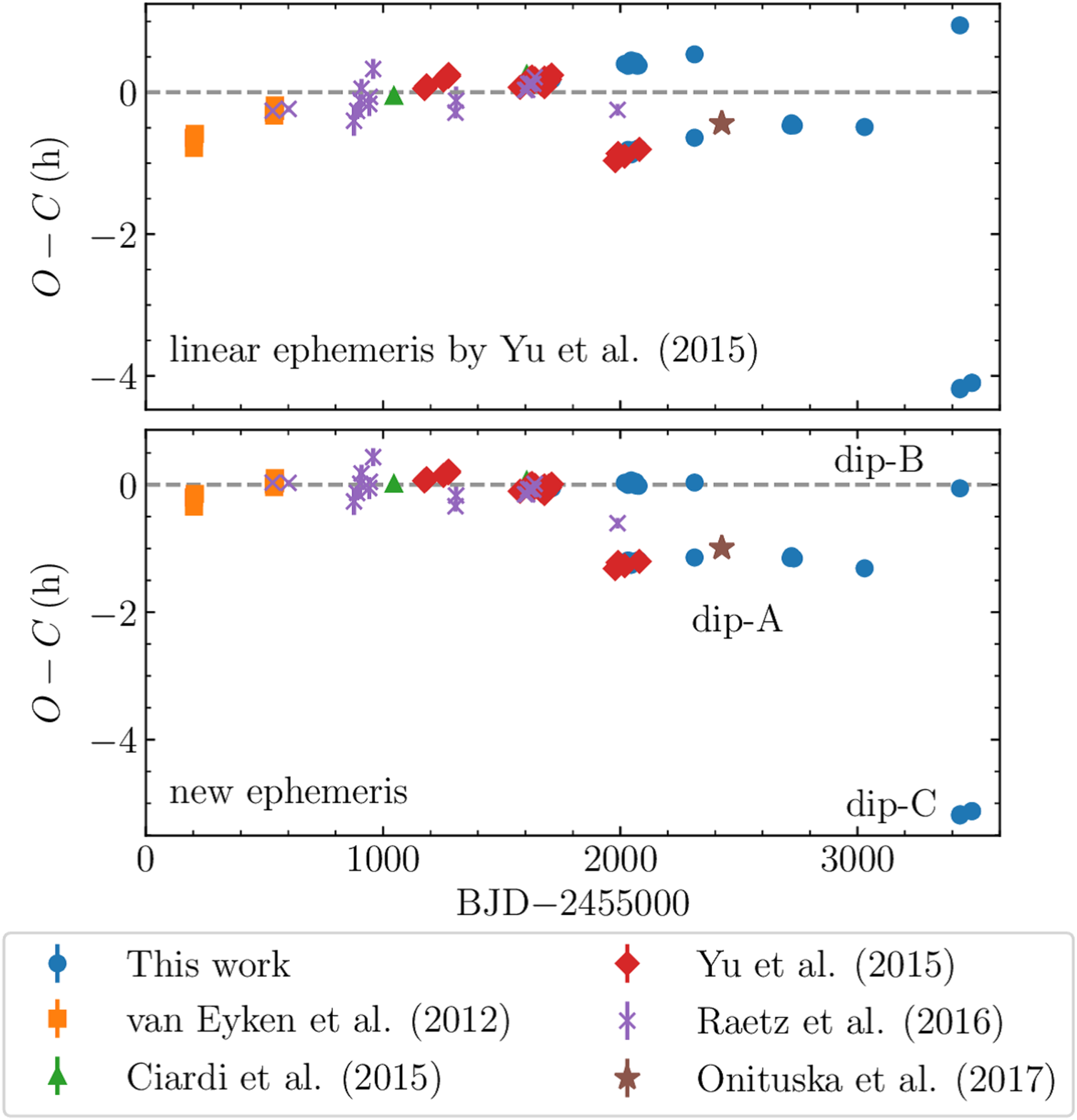} 
        \end{center}
      \caption{$O - C$ diagrams of all observed transit centers including the previous studies. Ephemeris proposed by \citet{Yu2015} and us are employed to plot the top and bottom panels, respectively.}\label{fig:O-C}
      \end{figure}

      We drew an ``Observed minus Calculated'' ($O - C$) diagram for all fading events using the ephemeris reported in \citet{Yu2015} as the top panel of figure~\ref{fig:O-C}, for revealing the periodicity of these mysterious events.
      We referred to table~4 in \citet{Ciardi2015}, table~1 in \citet{Yu2015}, table~B1 in \citet{Raetz2016}, and table~1 in \citet{Onitsuka2017} to plot.
      Note that the data observed by \citet{VanEyken2012} were taken from table~1 in \citet{Yu2015} and only ``complete'' data in \citet{Raetz2016} were employed.

      In the top panel, there is a sequence from data observed by \citet{VanEyken2012} to dip-B.
      Then, assuming that they have the same source and dip-A has a different origin from dip-B, new ephemeris was derived using only dip-B in our data.
      We fitted the center times of dip-B listed in table~\ref{tab:allparams} by weighted least squares method.
      The best-fitting parameters are
      \begin{equation}\label{eq:ephemeris}
        \begin{array}{rcl}
          T_0\>\mathrm{[BJD_{TDB}]} & = & 2455543.943\,\pm\,0.002, \\
          P & = & 0.4483993\,\pm\,0.0000006\>\mathrm{d}.
        \end{array}
      \end{equation}
      Our new $O - C$ diagram shown in the lower panel of figure~\ref{fig:O-C} indicates that this ephemeris is also consistent with the previously reported events.
      Note that the difference between our ephemeris and that proposed by \citet{Yu2015} originates from the way of dealing with dip-A.
      We consider that dip-A reflects a phenomenon different from dip-B, while they recognized the two dips as single fading events.
      This is why they interpreted that the times for the dip center had shifted from the extrapolation of the past events.
      Although the authors do not mentioned, one can see a deeper dip at around phase -0.1 and a shallower dip at around phase 0 in the 2014 November 29 and 2014 December 27 data in figure 3 of \citet{Yu2015}, and in the bottom panel of figures 2, 3, and 4 of \citet{Koen2015}.
      Note that the predicted times of planetary transit (phase 0) is shifted earlier by 0.035 days based on our new ephemeris in the figures of \citet{Koen2015}.
      These data are consistent with our detection of dip-A and B.

    \subsection{Wavelength dependence}
\begin{figure*}
        \begin{center}
          \includegraphics[width=16cm]{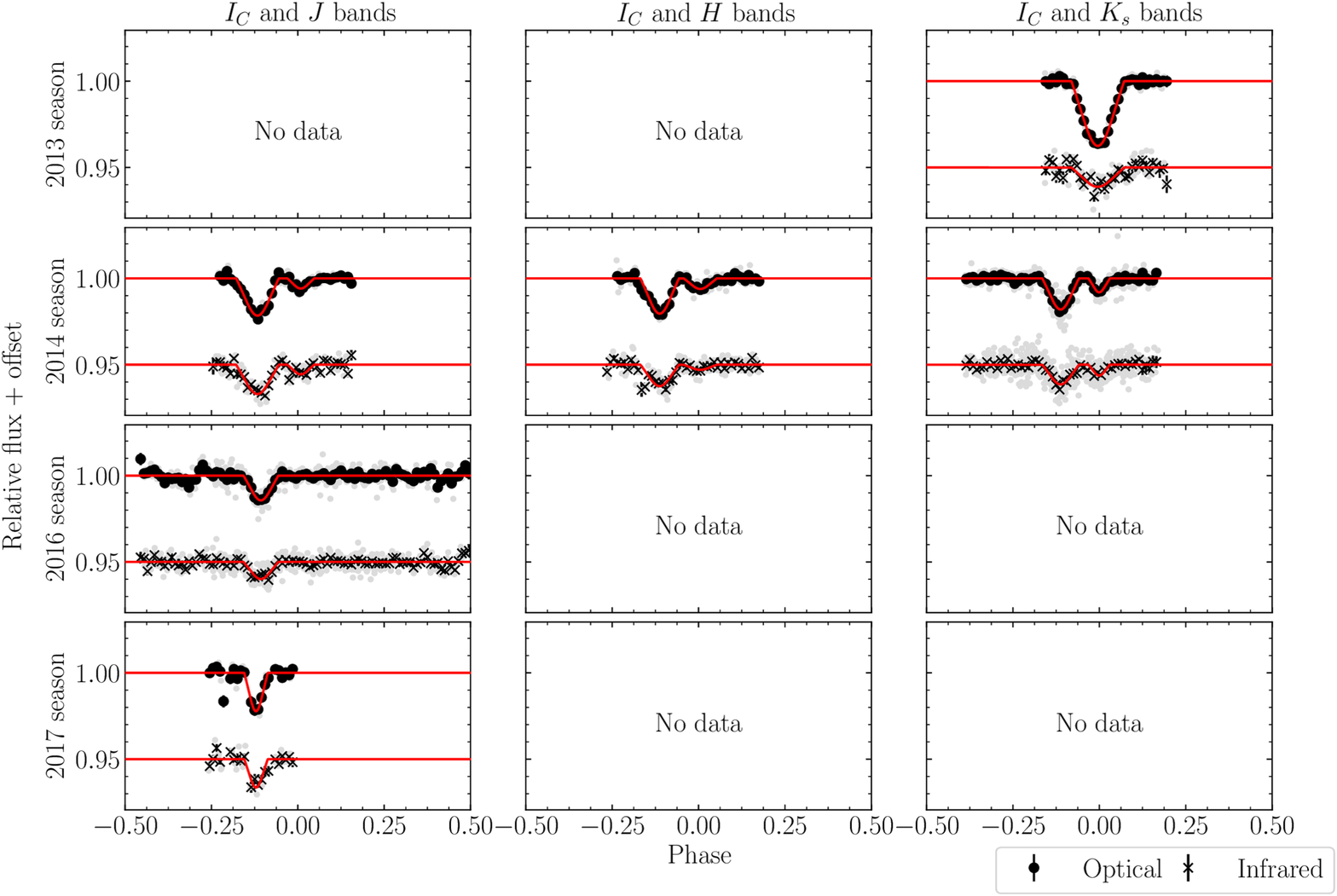}
        \end{center}
        \caption{Phase-folded light curves aligned according to their observational seasons and infrared filters at the time. Panels on the same row are data obtained in the same season, while those on the same column are data obtained by the same combination of optical and infrared filters. Gray dots indicate raw phase-folded light curves. Black circles and crosses indicate optical and infrared data binned by 0.01 in phase, respectively. Solid lines are the best-fitting dip functions.}\label{fig:folded}
      \end{figure*}

      To reinvestigate the nature of this system, we examined the wavelength dependence of the depth.
      After flattening the light curves as shown in figure~\ref{fig:lc_def}, we conducted phase-folding within each observational season by using the new ephemeris and binning with a bin width of 0.01 in phase to increase the apparent \textit{S/N} ratios.
      To compare the optical data with the simultaneously observed infrared data, the optical data were grouped according to the infrared filter used at the time, and therefore, HOWPol data were excluded in the procedure.
      Next, we fitted these phase-folded light curves with the dip function in equation~(\ref{eq:f_tr}) in the same way as the initial fitting, except that the trend function was fixed (see subsection~\ref{sec:fitting}).
      All phase-folded light curves and the best-fitting fading depths are shown in figure~\ref{fig:folded} and table~\ref{tab:wldepend}, respectively.
      Note that the data obtained on 2018 February 8 were excluded in this step because no fading events were detected.
      Therefore, the 2017 season light curves are identical to those on 2017 October 3.
      \begin{table*}
        \tbl{The best-fitting dip parameters for phase-folded and binned light curves.}{
        \begin{tabular}{lccccc} \hline
          Season & Infrared & Depth in the optical & Depth in the infrared & Ratio of depth & Duration \\
          & band & $\delta_\mathrm{opt}\>(\%)$ & $\delta_\mathrm{ira}\>(\%)$ & $\delta_\mathrm{ira} / \delta_\mathrm{opt}$ & $w\>$(in phase) \\ \hline
          \multicolumn{6}{c}{dip-A}  \\
          2014 & \textit{J} & $2.16\,\pm\,0.07$ & $1.70\,\pm\,0.12$ & $0.79\,\pm\, 0.06$ & $0.124\,\pm\, 0.004$ \\
          2014 & \textit{H} & $2.04\,\pm\,0.06$ & $1.25\,\pm\,0.10$ & $0.61\,\pm\, 0.05$ & $0.112\,\pm\, 0.004$ \\
          2014 & $K_s$ & $1.80\,\pm\,0.05$ & $1.14\,\pm\,0.07$ & $0.63\,\pm\, 0.04$ & $0.106\,\pm\, 0.002$ \\
          2016 & \textit{J} & $1.37\,\pm\,0.07$ & $0.92\,\pm\,0.07$ & $0.67\,\pm\, 0.06$ & $0.101\,\pm\, 0.004$ \\
          2017 & \textit{J} & $2.23\,\pm\,0.18$ & $1.64\,\pm\,0.16$ & $0.74\,\pm\, 0.10$ & $0.068\,\pm\, 0.004$ \\
          \hline
          \multicolumn{6}{c}{dip-B}  \\
          2014 & \textit{J} & $0.59\,\pm\,0.08$ & $0.56\,\pm\,0.14$ & $0.95\,\pm\, 0.27$ & $0.078\,\pm\, 0.011$ \\
          2014 & \textit{H} & $0.62\,\pm\,0.06$ & $0.30\,\pm\,0.11$ & $0.49\,\pm\, 0.18$ & $0.096\,\pm\, 0.012$ \\
          2014 & $K_s$ & $0.81\,\pm\,0.06$ & $0.63\,\pm\,0.09$ & $0.78\,\pm\, 0.12$ & $0.062\,\pm\, 0.004$ \\
          \hline
          \multicolumn{6}{c}{dip-C}  \\
          2018 & \textit{J} & $1.30\,\pm\,0.06$ & $0.95\,\pm\,0.09$ & $0.73\,\pm\, 0.07$ & $0.110\,\pm\, 0.004$ \\
          \hline
          \multicolumn{6}{c}{before splitting}  \\
          2013 & $K_s$ & $3.75\,\pm\,0.05$ & $1.12\,\pm\,0.12$ & $0.30\,\pm\, 0.03$ & $0.153\,\pm\, 0.003$ \\ \hline
        \end{tabular}}\label{tab:wldepend}    
      \end{table*}
      
      As listed in table~\ref{tab:wldepend}, both dip-A's and dip-C's infrared-to-optical depth ratios are about 0.7.
      Even if we consider their errors, the ratios do not include 1.
      In other words, the observed depths of dip-A and C depend on the wavelength.
      Note that this is consistent with the wavelength dependence reported in \citet{Onitsuka2017} because the fading event discussed by them is dip-A (see figure~\ref{fig:O-C}), which shows wavelength dependence in our data, too.
      In contrast, dip-B's ratios of infrared to $I_C$-band depth include 1 in $2\,\sigma$ range for \textit{J-} and $K_s$-band data, unlike dip-A and dip-C.
      This possible absence of wavelength dependence on dip-B is a new finding.
  
   \section{Discussion} \label{sec:discussion}
    As described above, we classified the newly detected fading events into three types based on their phases.
    Now, we examine four candidates for the source of dips-A, B, and C: a starspot, an accretion hotspot, a planet, and a dust clump.
    The fading event before splitting detected on 2014 February 23 is discussed later.
    
    \subsection{Origin of dip-A}
      \subsubsection{Cool starspot}\label{sec:coolspotA}
        WTTS are well-known for their large and long-lived starspots (e.g., \cite{Mahmud2011}) and such a starspot may produce a transit-like dip if it is located near the pole (e.g., \cite{Joergens2001}).
        Some of the previous studies inspected whether the cool starspot model is consistent with the observations, because the change in the shape and depth of the fading events could easily result from varying starspot distribution.
        However, all of them disfavored this hypothesis mainly because of the difficulty in producing both the observed duration and depth \citep{VanEyken2012, Ciardi2015, Yu2015}.

        Because these discussions were not about dip-A but about the single fading event before splitting, here we examine this hypothesis more quantitatively.
        We assume that the stellar surface has simply two temperature components (a stellar one, $T_*$, and a spot one, $T_\mathrm{cool}$) and that their brightness follows blackbody radiation.
        When the stellar surface is covered by a starspot with a filling factor $f$, dip-A's observed depth $\delta_\mathrm{cool}(\lambda)$ at wavelength $\lambda$ is represented by
        \begin{equation}\label{eq:depth_spot}
          \delta_\mathrm{cool}(\lambda) = \frac{f[B_\lambda(T_*) - B_\lambda(T_\mathrm{cool})]}{B_\lambda(T_*)}.
        \end{equation}
        Note that, for simplicity, we do not consider the limb-darkening effect.
        Hence, the ratio of depths observed at two different wavelengths is
        \begin{equation}
          \frac{\delta_\mathrm{cool}(\lambda_1)}{\delta_\mathrm{cool}(\lambda_2)} 
          = \frac{B_{\lambda_2}(T_\mathrm{cool})}{B_{\lambda_1}(T_\mathrm{cool})} \cdot 
          \frac{B_{\lambda_1}(T_*) - B_{\lambda_1}(T_\mathrm{cool})}{B_{\lambda_2}(T_*) - B_{\lambda_2}(T_\mathrm{cool})},
        \end{equation}
        which depends on only $T_*$ and $T_\mathrm{cool}$.
        Then, fixing $T_*$ to $3470\>\mathrm{K}$ \citep{Briceno2005}, we fit the obtained ratios shown in table~\ref{tab:wldepend}.
        The best-fitting spot temperatures are $T_\mathrm{cool} = 2740\,\pm\,110\>\mathrm{K}$ for the 2014 season and $T_\mathrm{cool} = 3400\,\pm\,360\>\mathrm{K}$ for the 2016 season.

\begin{figure}
          \begin{center}
            \includegraphics[width=8cm]{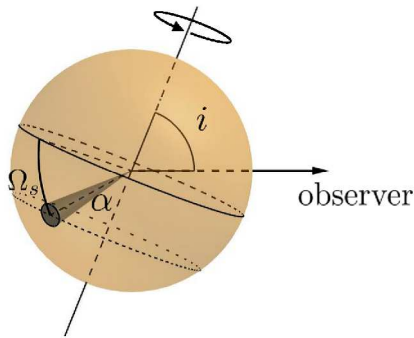}
          \end{center}
          \caption{Illustration of starspot's geometry. In this case, $\Omega_s$ has a negative value.}\label{fig:spot}
        \end{figure}

\begin{figure}
          \begin{center}
            \includegraphics[width=8cm]{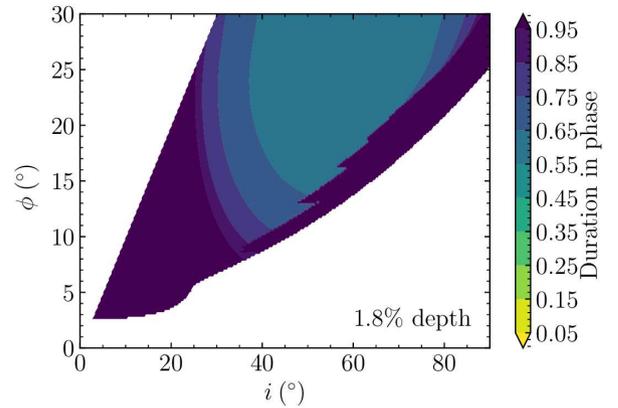}
          \end{center}
          \caption{Contour map of duration of the model starspot for dip-A. Spot temperature and size are adjusted to reproduce dip-A in the 2014 season. Spots in the white region cannot reproduce the observed depth.}\label{fig:spot_testA14}
        \end{figure}

\begin{figure}
          \begin{center}
            \includegraphics[width=8cm]{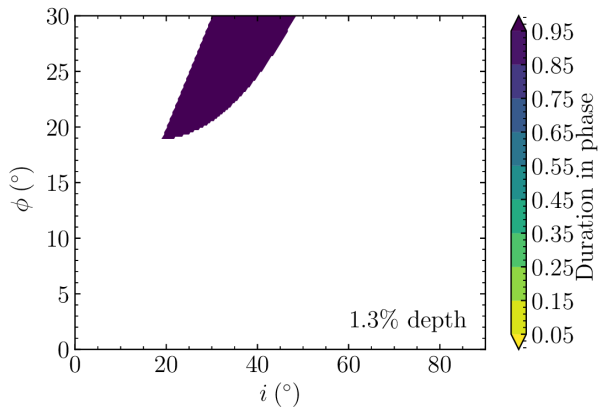}
          \end{center}
          \caption{Same as figure~\ref{fig:spot_testA14}, but for dip-A in the 2016 season.}\label{fig:spot_testA16}
        \end{figure}

        Next, we test whether a single starspot with these temperatures can reproduce the $I_C$-band light curves.
        We model a circular starspot with an angular radius $\alpha$ at latitude $\Omega_s (-90^\circ < \Omega_s < 90^\circ)$ on a star whose rotation axis is inclined at $i (0^\circ < i < 90^\circ)$ from the line of sight, as shown in figure~\ref{fig:spot}.
        Because the spot should pass the edge of the photosphere in order to mimic a transit-like dip, we employ an alternative parameter, $\phi \equiv i + \Omega_s$ (i.e., $\phi$ must be at most a few tens of degrees independent from $i$ when the spot grazes).
        After calculating $\alpha$ for each combination of $i$ and $\phi$, which satisfies the observed depths [i.e., 1.8\% for the 2014 season and 1.3\% for the 2016 season (see also table~\ref{tab:wldepend})], we derive the duration of the dip from $\alpha$, $i$, and $\phi$.
        The results of this simple simulation are shown in figures~\ref{fig:spot_testA14} and~\ref{fig:spot_testA16}.
        Note that we do not compute the case in which the spot covers the stellar pole (i.e., $\alpha$ has its upper limit for every combination of $i$ and $\Omega_s$).
        These figures show that starspots for reproducing dip-A in the 2014 and the 2016 season yield more than 0.45 and 0.85 durations in phase, respectively.
        Therefore, a starspot is an unlikely origin of dip-A.

        Furthermore, we examine whether limb-darkening affects these results.
        If employing the quadratic limb-darkening coefficients $(c_1, c_2) = (0.3840, 0.3447)$, which are calculated for $I_C$ band, $\log\,g = 3.5$ and $T_\mathrm{eff} = 3500\>\mathrm{K}$ \citep{Claret2013}, the integrated stellar flux becomes $\sim 0.8$ times weaker than the uniform surface brightness.
        This reduced stellar flux can yield at most 1.25 times deeper depth than the uniform case.
        However, there is still no parameter space to satisfy both the observed duration and 1.6\% (= 2.0\% / 1.25) depth.
        From these results, we conclude that the starspot is not the source of dip-A.

      \subsubsection{Hotspot}\label{sec:hotspotA}
        Accretion from a disk to a central star along a stellar magnetosphere shapes a hotspot at the stellar surface.
        Although this phenomenon is common not to WTTS but to classical T-Tauri stars \citep{Herbst1994}, a corotating hotspot at high latitude, if exists, can yield a transit-like dip.
        We test this hypothesis with the same assumption as that of the starspot model (see sub-subsection~\ref{sec:coolspotA}).
        The depth of a fading event yielded by a hotspot whose temperature is $T_\mathrm{hot}$ is represented by
        \begin{equation}
          \delta_\mathrm{hot}(\lambda) =
          \frac{f[B_\lambda(T_\mathrm{hot}) - B_\lambda(T_*)]}{(1 - f)B_\lambda(T_*) + fB_\lambda(T_\mathrm{hot})},
        \end{equation}
        where $f$ is a filling factor.
        Unlike the starspot model, two parameters (i.e., $T_\mathrm{hot}$ and $f$) are necessary to calculate the ratio of depths at two wavelengths.

\begin{figure}
          \begin{center}
            \includegraphics[width=8cm]{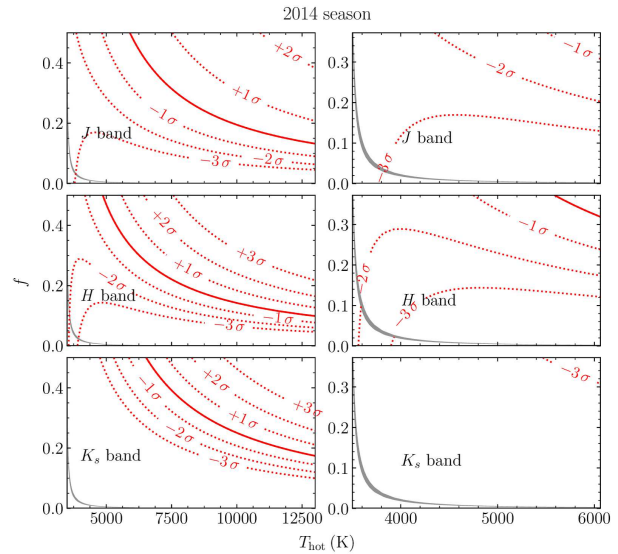}
          \end{center}
          \caption{Contour maps of the infrared-to-optical depth ratio of dip-A in the 2014 season. \textit{J-, H-} and $K_s$-band data are shown in the top, middle, and bottom panels, respectively. Panels on the right column are enlarged ones of those on the left column. Solid lines indicate loci of the infrared-to-optical depth ratios: 0.79 for the \textit{J} band, 0.61 for the \textit{H} band, and 0.63 for the $K_s$ band. Dashed lines indicate a few $\sigma$ levels of the observed depth ratio (see also table~\ref{tab:wldepend}). Gray areas are parameter spaces to yield 1.75\%--2.25\% depth at $I_C$ band.}\label{fig:hotspot_A14}
        \end{figure}

\begin{figure}
          \begin{center}
            \includegraphics[width=8cm]{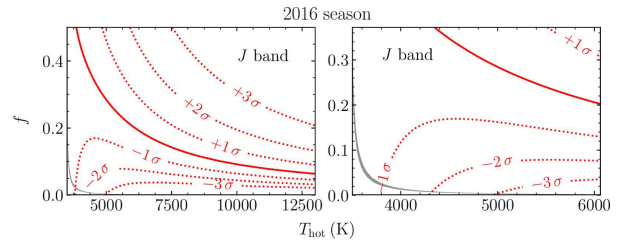}
          \end{center}
          \caption{Same as figure~\ref{fig:hotspot_A14} but for the 2016 season. Gray areas are parameter spaces to yield 1.2\%--1.5\% depth in $I_C$ band.}\label{fig:hotspot_A16}
        \end{figure}

        We draw contour maps of the 2014 and 2016 season's infrared-to-optical depth ratios for combinations of $T_\mathrm{hot}$ and $f$, as shown in figures~\ref{fig:hotspot_A14} and~\ref{fig:hotspot_A16} for the 2014 and 2016 seasons, respectively.
        To reproduce the observed typical depths at $I_C$ band, 1.75\%--2.25\% for the 2014 season and 1.2\%--1.5\% for the 2016 season, the parameters in gray areas at the lower left corners of these figures are necessary.
        Allowing $3\,\sigma$ range for the infrared-to-optical depth ratios, a hotspot at lower than about $3800\>\mathrm{K}$ can satisfy the \textit{J-} and \textit{H-}band ratios for the 2014 season (see the top and middle panels on the right side of figure~\ref{fig:hotspot_A14}).
        This temperature is also consistent with the \textit{J-}band ratio for the 2016 season.
        However, such a cool temperature is inappropriate for the name of hotspot because the temperature is only $300\>\mathrm{K}$ higher than that of the stellar surface, $3500\>\mathrm{K}$.
        Furthermore, even lower temperature and a very large filling factor, $f > 0.5$, is needed for the $K_s$-band ratio for the 2014 season (see the bottom figures of figure~\ref{fig:hotspot_A14}).
        Thus, we do not support the hotspot hypothesis.

        \subsubsection{Planet}
          From previous discussions, dip-A is not caused by a phenomenon at the stellar surface.
          Therefore, a transiting planet is an alternative hypothesis.
          Strictly speaking, the planetary radius shows a slight wavelength dependence due to its atmosphere; however, such dependence would be negligible considering our low \textit{S/N} level.
          As listed in table~\ref{tab:wldepend}, $3\,\sigma$ ranges of infrared-to-optical depth ratios in both the 2014 and 2016 seasons do not allow unity, which would be expected for a planetary transit.
          Therefore, we can rule out this hypothesis for dip-A in the 2014 and 2016 seasons.

        \subsubsection{Dust clump}\label{sec:dust_A}
\begin{figure}
            \begin{center}
              \includegraphics[width=8cm]{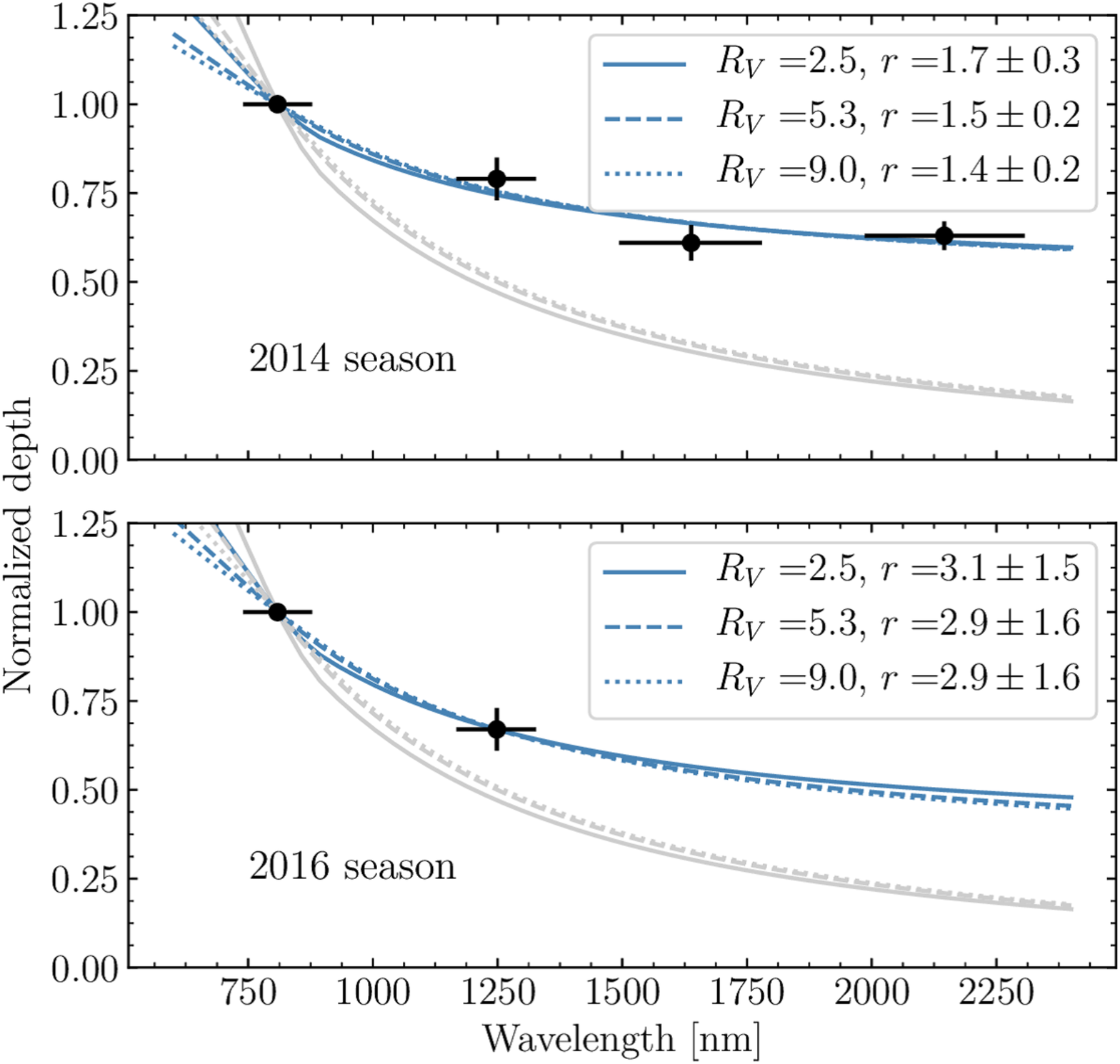}        
            \end{center}
            \caption{Wavelength dependence of the normalized dip depth of dip-A in the 2014 season (top panel) and the 2016 season (bottom panel). Gray lines indicate the estimated wavelength dependence from an optically thin dust cloud, while the blue ones indicate the best-fitting core-halo dust clump models for each value of $R_V$. The best-fitting values of $r$ are shown in the legends.}\label{fig:dust_A}
          \end{figure}

          The last possibility is the dust clump hypothesis, which is concluded as the origin of dip-A by \citet{Onitsuka2017}.
          Young stars with quasi-periodic fading events are usually called ``dipper'' stars (e.g., \cite{Ansdell2015}).
          With rather deep depths as few tens of percents, such fading events are thought to be caused by a circumstellar dust clump \citep{David2017}.
          To begin with, we examine if an optically thin dust cloud can reproduce the observed wavelength dependence.
          Assuming that the optically thin dust cloud (i.e., optical depth $\tau \ll 1$) follows the extinction law proposed by \citet{Cardelli1989}, we represent the depth caused by the dust cloud as 
          \begin{equation}\label{eq:d_dust}
            \delta_\mathrm{dust}(\lambda) = f_\mathrm{dust}\tau(\lambda) = f_\mathrm{dust}\tau_V\left[a(1/\lambda)+\frac{b(1/\lambda)}{R_V}\right],
          \end{equation}
          where $f_\mathrm{dust}$ is a filling factor of the dust clump, $\tau_V$ is the optical depth at \textit{V} band, $a(1/\lambda)$ and $b(1/\lambda)$ are the wavelength-dependent coefficients reported in \citet{Cardelli1989}, and $R_V$ is a ratio of total to selective extinction ($\equiv A_V / E(B-V)$).
          Thus, the infrared-to-optical depth ratios depend on only $R_V$.
          However, this estimated wavelength dependence is not consistent with our observations, as the gray lines in figure~\ref{fig:dust_A}.

          For this model to satisfy the wavelength dependence, we assume that this clump consists of two components: a completely opaque core and an optically thin dust halo, which follows the interstellar dust extinction.
          Note that the core might not be dust but a planet because it should only be opaque.
          The depth yielded by the optically thin component is represented by equation~(\ref{eq:d_dust}).
          On the other hand, the depth yielded by the core is equal to its filling factor $f_\mathrm{core}$ independent of the wavelength.
          Because the observed depth is a sum of these two depths, as
          \begin{equation}
            \delta_\mathrm{obs}(\lambda)
            = f_\mathrm{core} + f_\mathrm{halo}\tau_V\left[a(1/\lambda)+\frac{b(1/\lambda)}{R_V}\right],
          \end{equation}
          a ratio of depths observed in two different wavelengths depends on two parameters, $r \equiv f_\mathrm{halo}\tau_V / f_\mathrm{core}$ and $R_V$.
          Then, we search the best-fitting $r$ for certain values of $R_V$, as shown in figure~\ref{fig:dust_A}.

          Fixing $r$ to the best-fitting value at $R_V = 5.3$ because the fitting results do not depend on the value of $R_V$ much, we derive $f_\mathrm{core}$ and $f_\mathrm{halo}\tau_V$ from the observed dip depth at $I_C$ band, as listed in table~\ref{tab:dust}.
          \begin{table}
            \tbl{Best-fitting dust-clump parameters for dip-A}{
            \begin{tabular}{lcccc} \hline
              Season & Typical depth in optical$\>(\%)$ & $f_\mathrm{core}$ & $\sqrt{f_\mathrm{core}}$ & $f_\mathrm{halo}\tau_V$ \\ \hline
              2014 & 2.0 & 0.01  & 0.1 & 0.014 \\
              2016 & 1.5 & 0.005 & 0.07 & 0.0135 \\ \hline
            \end{tabular}}\label{tab:dust}
          \end{table}
          Although we cannot claim more quantitatively, we emphasize that these obtained values are consistent with the optically thin assumption.
          If these values are correct, the core-to-star radius ratios, $\sqrt{f_\mathrm{core}}$, are 0.1 and 0.07 in the 2014 and 2016 seasons, respectively.
          Comparing the best-fitting values in the two seasons, $f_\mathrm{core}$ in the 2014 season is twice that in the 2016 season, whereas $f_\mathrm{halo}\tau_V$ in the 2014 season is slightly smaller than that in the 2016 season.
          This indicates that the core may be collapsing into the dust cloud.

        \subsection{Origin of dip-B}
          \subsubsection{Cool starspot}
            We examine this hypothesis as sub-subsection~\ref{sec:coolspotA}.
            The derived spot temperature from the wavelength dependence of dip-B is $T_\mathrm{cool} = 2410\,\pm\,340\>\mathrm{K}$.
            Then, fixing the spot temperature to $2410\>\mathrm{K}$, we examine whether this spot satisfies the observed depth and duration, as listed in table~\ref{tab:wldepend}.
            \begin{figure}
              \begin{center}
                \includegraphics[width=8cm]{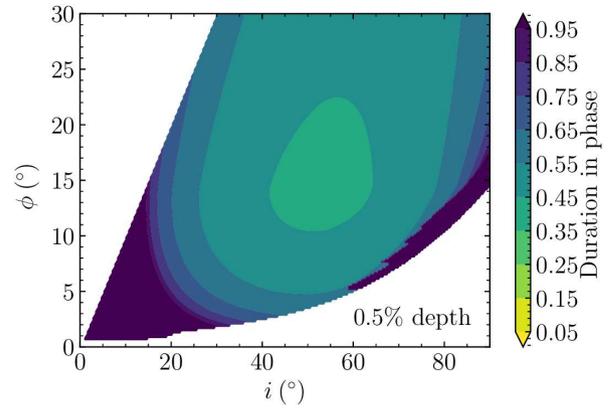}
              \end{center}
              \caption{Same as figure~\ref{fig:spot_testA14}, but for dip-B.}\label{fig:spot_test} 
            \end{figure}
            As shown in figure~\ref{fig:spot_test}, all spots that reproduce 0.5\% depth would yield durations longer than 0.35 in phase, which is much larger than the observed values of 0.08.
            In addition, although we tested the limb-darkened case, spot yield 0.4\% (= 0.5\% / 1.25) depth could not satisfy the observed duration, either.
            Hence, this hypothesis is ruled out for dip-B.

          \subsubsection{Hotspot}\label{sec:hotspot}
\begin{figure}
              \begin{center}
                \includegraphics[width=8cm]{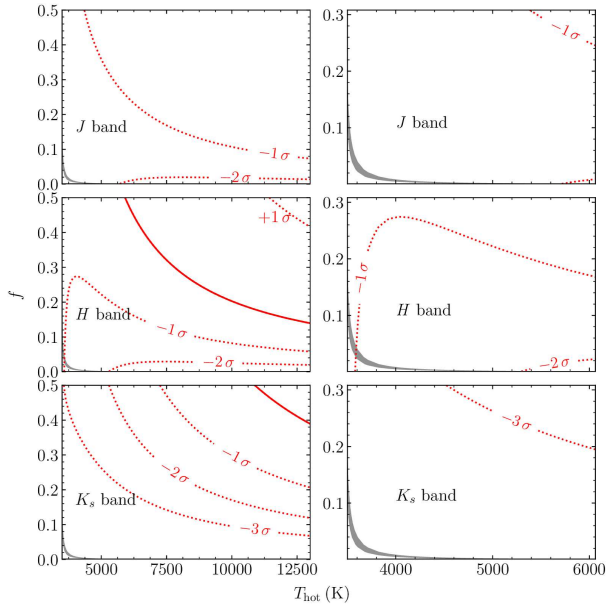}                
              \end{center}
              \caption{Same as figure~\ref{fig:hotspot_A14}, but for dip-B.}\label{fig:hotspot}
            \end{figure}

            In the same way as sub-subsection~\ref{sec:hotspotA}, we tested whether it is possible to reproduce both the observed depth at $I_C$ band and the wavelength dependence.
            Figure~\ref{fig:hotspot} is the result of this test.
            Although \textit{J-} and \textit{H-}band depth ratios yielded by parameters in the gray areas are consistent with the observed ones in $2\,\sigma$ level, the $K_s$-band depth ratio is not consistent with $3\,\sigma$ level, as shown in figure~\ref{fig:hotspot}.
            Thus, we reject this hypothesis.

          \subsubsection{Planet}\label{sec:planet}
            Different from dip-A, dip-B's infrared-to-optical depth ratios are consistent with unity if $3\,\sigma$ errors are allowed. 
            This means that the planetary hypothesis cannot be denied based on the wavelength dependence.
            Although one difficulty with this hypothesis is the variation in depths and the disappearance of dip-B, the spin-orbit nodal precession can solve this problem.
      
            We extract profiles of dip-B from the optical light curves observed from 2014 December 27 to 2015 October 17 when dip-B is detected.
            Then, we fit these profiles with a numerical transit light curve model, \verb|batman| \citep{Kreidberg2015}, assuming that the planetary orbit is constant.
            Fixing the orbital eccentricity to zero and the quadratic limb-darkening coefficients to $(c_1, c_2) = (0.3840, 0.3447)$, we derive the orbital semi-major axis in units of stellar radius $a/R_*$, ratio of the planet radius to the stellar radius $R_p / R_*$, and inclination angle $i$ ($i = 90^\circ$ at an edge-on orbit) by using the Markov Chain Monte Carlo (MCMC) method.
            The best-fitting parameters from $10^7\>$MCMC steps, except the first $10^6\>$steps, are listed in table~\ref{tab:orb_param}.
            We discuss the derived planetary parameters later.
            \begin{table}
              \tbl{List of planetary parameters.}{
                \begin{tabular}{c@{\hspace{1.5cm}}c}
                  \hline
                  Parameter & Value \\ \hline
                  \multicolumn{2}{c}{Measured in subsection~\ref{sec:ephemeris}} \\
                  $P_\mathrm{orb}\>\mathrm{(d)}$ & $0.4483993\,\pm\,0.0000006$ \\ \hline
                  \multicolumn{2}{c}{Measured from 2014 sesason} \\
                  $a / R_*$ & $3.94\,\pm\,1.55$\footnotemark[$*$] \\
                  $R_p / R_*$ & $0.083\,\pm\,0.011$\footnotemark[$*$] \\
                  $i\>({}^\circ)$ & $78.4\,\pm\,11.5$\footnotemark[$*$] \\ \hline
                  \multicolumn{2}{c}{Derived} \\
                  $a$ & $1.79\,\pm\,0.15\,R_{\solar}$\footnotemark[$\dagger$] \\
                  $R_*$ & $0.45\,\pm\,0.18\,R_{\solar}$ \\
                  $R_p$ & $0.36\,\pm\,0.15\,R_\mathrm{Jup}$ \\ \hline
                \end{tabular}}\label{tab:orb_param}
                \begin{tabnote}
                  \footnotemark[*] Median value and half range of 68\% credibility interval. \\
                  \footnotemark[$\dagger$] Derived from Kepler's third law assuming $M_* = 0.39\,\pm\,0.10\,M_{\solar}$ following \citet{VanEyken2012} 
                \end{tabnote}
            \end{table}

\begin{figure}
              \begin{center}
                \includegraphics[width=8cm]{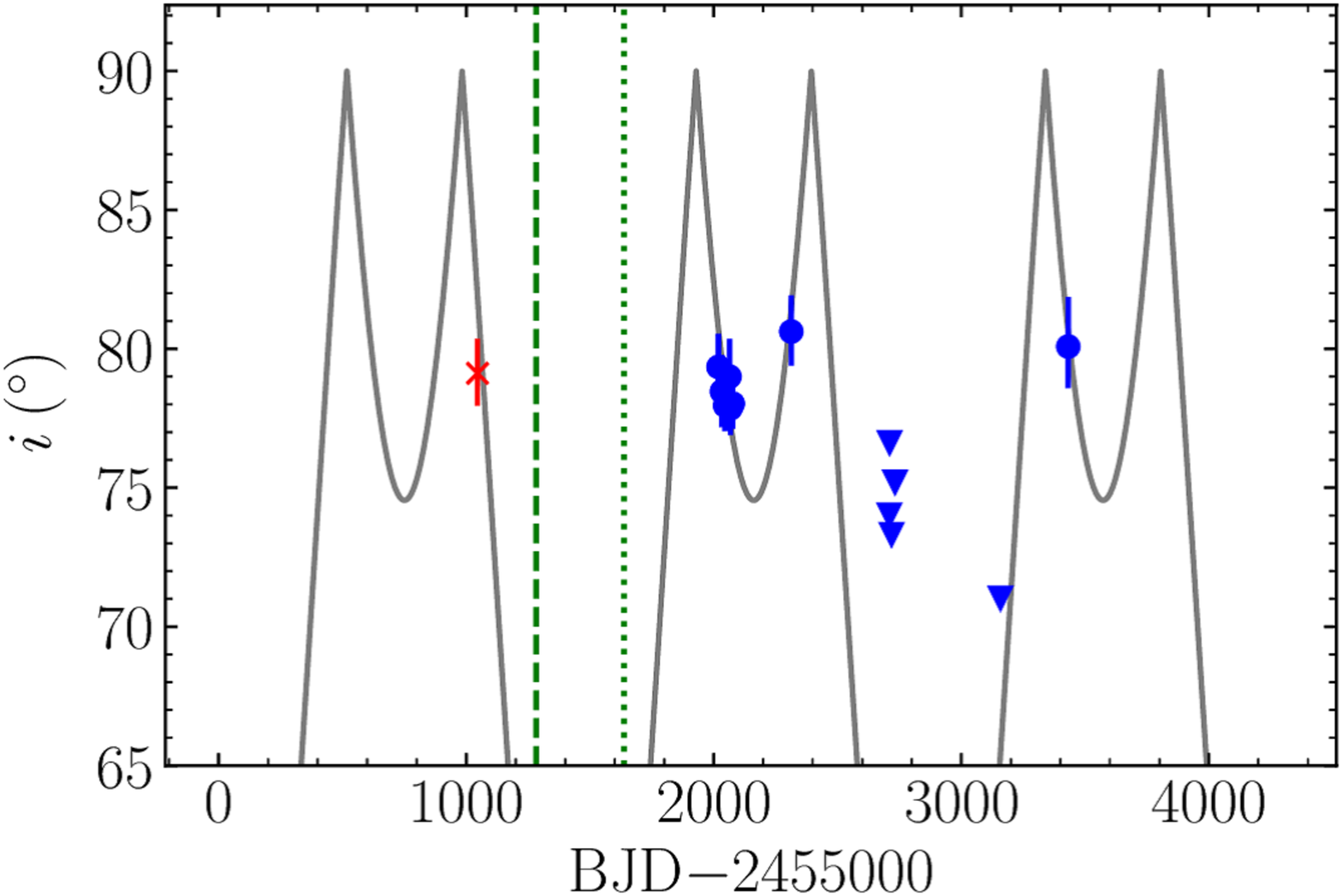}
              \end{center}
              \caption{Time variation of inclination angles. Circles and inverted triangles indicate the derived values and upper limits, respectively. The cross is derived from Spitzer data observed by \citet{Ciardi2015}. The solid line is the fitted sinusoidal function. The vertical dashed and dotted lines indicate the non-detections of RM effect in \citet{Ciardi2015} and \citet{Yu2015}, respectively.}\label{fig:inc}
            \end{figure}

            Then, fixing all parameters to the values listed in table~\ref{tab:orb_param}, except the inclination angle, we fit the inclination angle for each observational date individually.
            We derive inclination angles from $5 \times 10^6\>$MCMC steps, except for the first $5 \times 10^5\>$steps.
            Note that we set an upper limit for the inclination angle corresponding to the upper side of 95\% credibility interval when dip-B is not detected (see also table~\ref{tab:allparams}).
            Finally, we fit the derived inclination angles with a sinusoidal function replicated at $90^\circ$, as shown in figure~\ref{fig:inc}.
            The derived sinusoidal function is
            \begin{equation}
              i(t) =  31.5^\circ \sin \left(\frac{2\pi}{1410.9}(t - 245680.90)\right) + 73.9^\circ,
            \end{equation}
            where $t$ has unit of $\mathrm{BJD_{TDB}}$.
            This suggests that the observed variation in depths can be naturally explained through spin-orbit precession, whose period is $\sim 1411\>\mathrm{d}$.
            Note that this period is different from the previously proposed ones with gravity darkening \citep{Barnes2013,Kamiaka2015}.
            This is because the data used to derive the precession parameters are different.
            They used the data of dips before the splitting into dip-A and B, which is mostly dominated by the previous dip-A.
            On the other hand, we used only dip-B data after the splitting.

            One concern is that our derived parameters listed in table~\ref{tab:orb_param} imply a smaller stellar radius, $0.45\,\pm\,0.18\,R_{\solar}$ than that derived from the stellar luminosity in \citet{Briceno2005}.
            However, considering our large uncertainties in these parameters, this problem is not critical to deny this hypothesis now.

          \subsubsection{Dust clump}
\begin{figure}
              \begin{center}
                \includegraphics[width=8cm]{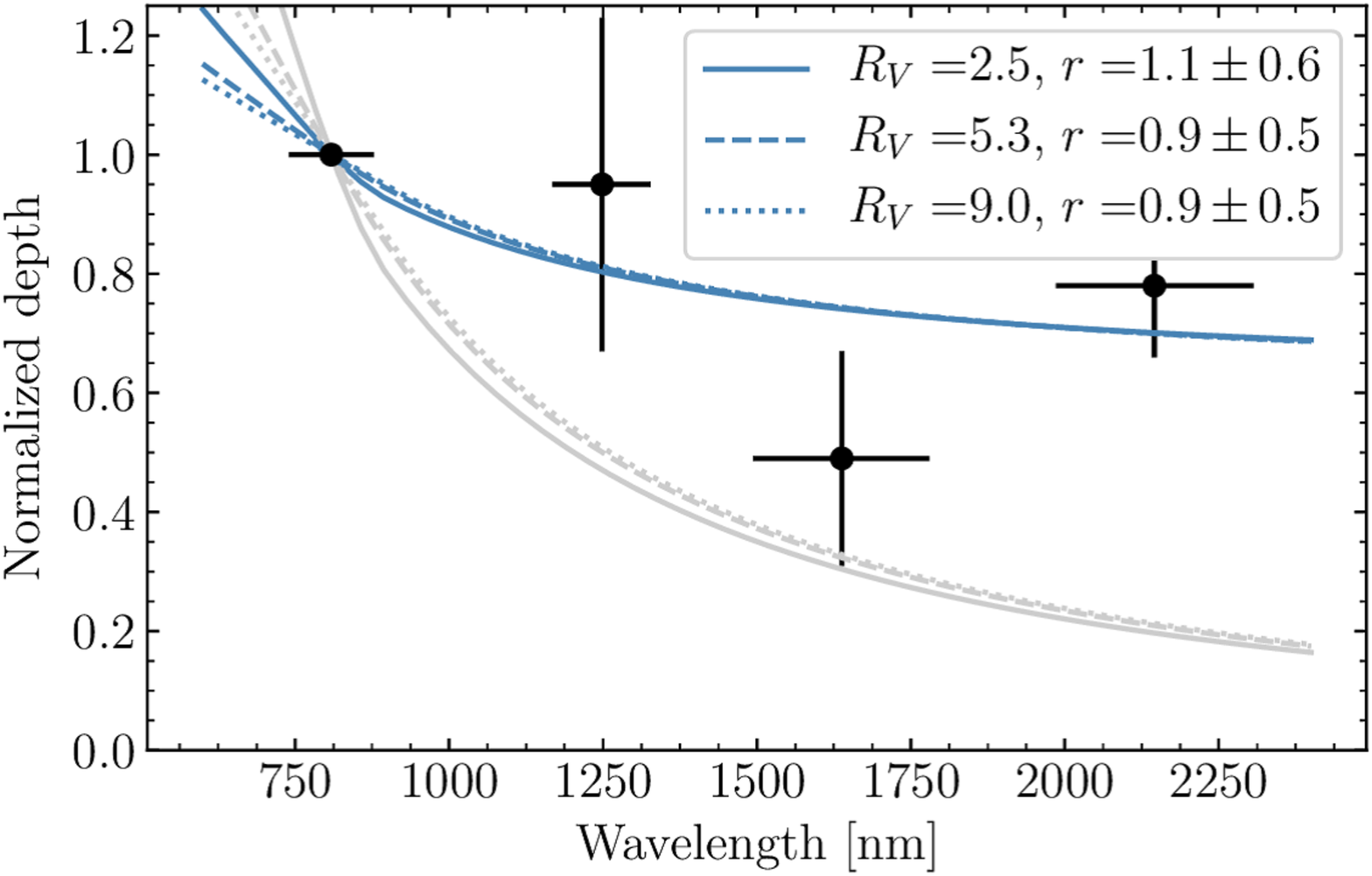}                
              \end{center}
              \caption{Same as figure~\ref{fig:dust_A}, but for dip-B.}\label{fig:dust_B}
            \end{figure}

            Although we examined the planetary hypothesis above based on the interpretation that dip-B has no wavelength dependence of the depth, here we test the dust clump hypothesis because the data are more consistent with the wavelength-dependent dip depth.
            Because an optically thin dust clump cannot satisfy the observed wavelength dependence, as shown in figure~\ref{fig:dust_B}, we fit the core-halo model as shown in sub-subsection~\ref{sec:dust_A}.
            The typical value of the best-fitting $r$ is 0.9 (see figure~\ref{fig:dust_B}).
            We derive $f_\mathrm{core}=0.003$ and $f_\mathrm{halo}\tau_V=0.0027$ from this value of $r$.
            As the case of dip-A, these values are not implausible.

            One problem with this hypothesis is the reappearance of dip-B.
            Dip-B disappeared after 2015 October 17, and reappeared on 2018 November 9.
            However, this hypothesis could overcome the difficulty, if this core is the precessing planet as discussed in the previous sub-subsection.
            In this case, the ratio of planetary radius to stellar radius is $\sqrt{f_\mathrm{core}}$, 0.05, which yields a planetary radius of $0.22\,R_\mathrm{Jup}$ in combination with the stellar radius of $0.45\,R_{\solar}$ from table~\ref{tab:orb_param}.
            Strictly speaking, this derivation is not correct because the stellar radius and ``the ratio of planetary radius to stellar radius'' are not solved simultaneously.
            The core-halo dust clump hypothesis assumes the whole dust clump passes over the stellar disk, i.e., near edge-on geometry.
            It is consistent with the inclination angle of $78.4^\circ$ in table~\ref{tab:orb_param} but a different light curve
            expected for the core-halo structure may require smaller inclination angle.
            Therefore, the derived planetary radius of $0.22\,R_\mathrm{Jup}$ may have a large uncertainty but is not a bad estimate for the putative planet embedded within the halo, i.e., the planetary radius can be considerably smaller than $0.36\,R_\mathrm{Jup}$.

        \subsection{Origin of dip-C}
          Dip-C appeared at around phase 0.5 in 2018 November and is recognized by us for the first time.
          Although the phase of dip-C is located at around 0.5, dip-C could not be a secondary eclipse because the planet should have a similar surface temperature to the host star in order to yield a similar depth as the primary transit.
          Then we discuss the four hypotheses for the source of dip-C as previous sections.

          \subsubsection{Cool starspot}
\begin{figure}
              \begin{center}
                \includegraphics[width=8cm]{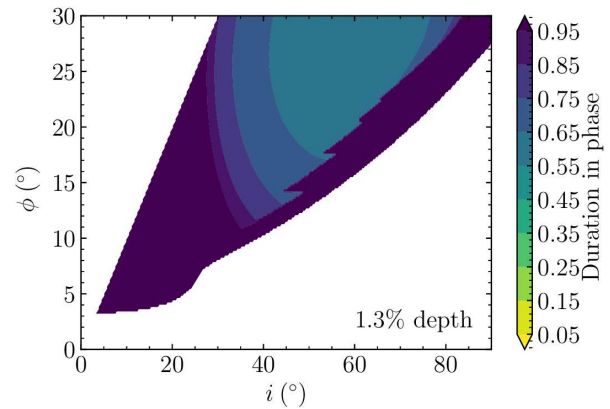}
              \end{center}
              \caption{Same as figure~\ref{fig:spot_testA14}, but for dip-C.}\label{fig:spot_testC}
            \end{figure}

            We derive the spot temperature from the infrared-to-optical depth ratio.
            The fitting result is $T_\mathrm{cool} = 3060\,\pm\,370\>\mathrm{K}$.
            Using this value, we test whether the observed depth at $I_C$ band, 1.3\%, is compatible with its duration, 0.11, in phase in the same way as that shown in sub-subsection~\ref{sec:coolspotA}.
            Figure~\ref{fig:spot_testC} shows the result of this test.
            This suggests that a single starspot yielding 1.3\% depth at $I_C$ band could not explain the observed short duration, and therefore, this hypothesis is ruled out.
            
          \subsubsection{Hotspot}
\begin{figure}
              \begin{center}
                \includegraphics[width=8cm]{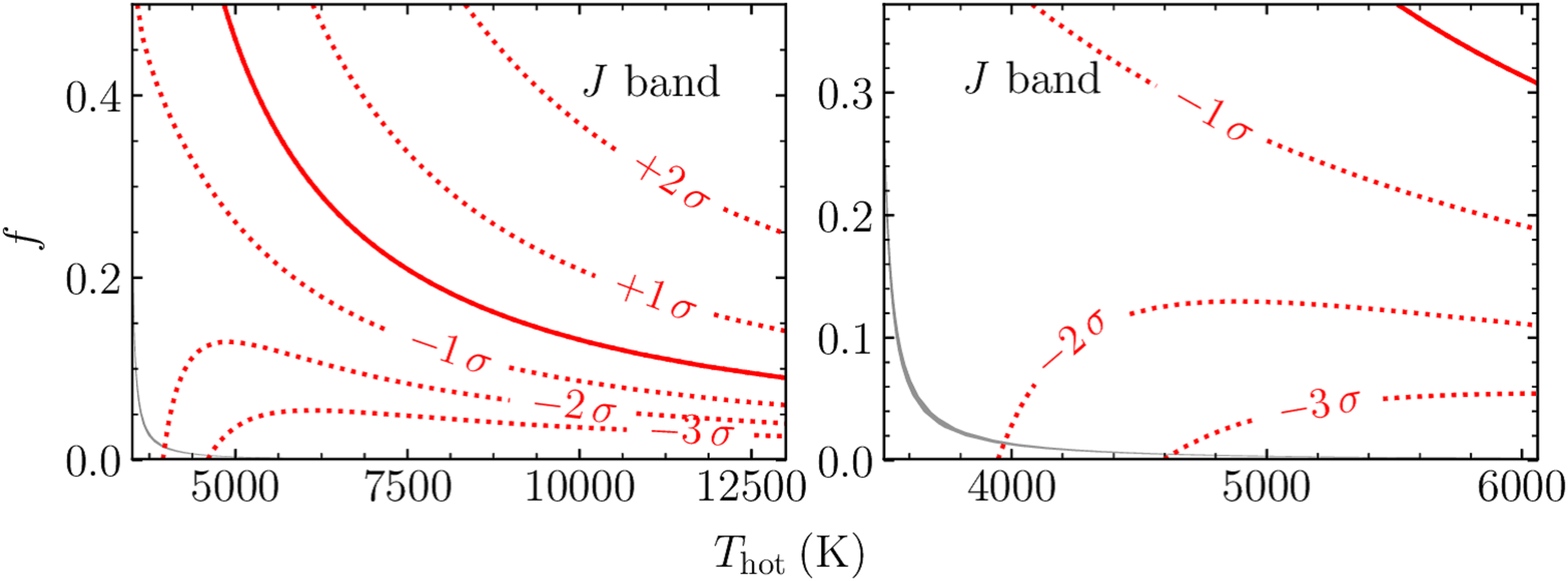}
              \end{center}
              \caption{Same as figure~\ref{fig:hotspot_A14}, but for dip-C.}\label{fig:hotspot_C}
            \end{figure}

          The result of the same inspection as sub-subsection~\ref{sec:hotspotA} is shown in figure~\ref{fig:hotspot_C}, and we could not rule out this hypothesis because the gray shaded area, which satisfies the observed depth at $I_C$ band, overlaps with $3\,\sigma$ range of the infrared-to-optical depth ratio.
            Therefore, a hotspot can reproduce the observed dip-C.
          
          \subsubsection{Planet}
            As listed in table~\ref{tab:wldepend}, the \textit{J-}to-$I_C$ depth ratio does not include 1 within $3\,\sigma$ range.
            Therefore, the planetary hypothesis is unlikely.
          
          \subsubsection{Dust clump}
\begin{figure}
              \begin{center}
                \includegraphics[width=8cm]{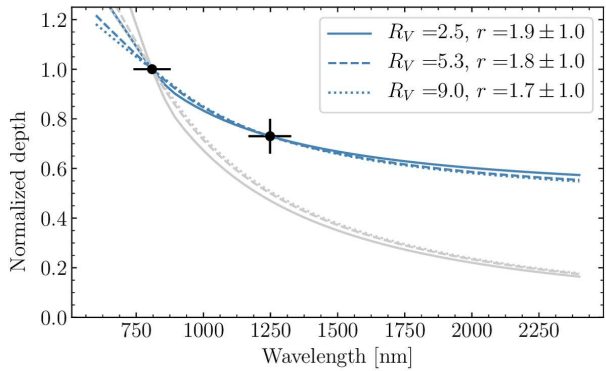}                
              \end{center}
              \caption{Same as figure~\ref{fig:dust_A}, but for dip-C.}\label{fig:dust_C}
            \end{figure}

            We estimate the wavelength dependence expected from the single-component dust clump model, as shown in figure~\ref{fig:dust_C}.
            Because this model is not adequate to satisfy the observational results, we fit the core-halo dust clump model and derive $r$ in the same way as that in sub-subsection~\ref{sec:dust_A}.
            The result of the fitting is shown in figure~\ref{fig:dust_C} and the best-fitting value is $r=1.4$ in $R_V=5.3$ case.
            To satisfy the depth at $I_C$ band at about 1.3\%, $f_\mathrm{core}$ and $f_\mathrm{halo}\tau_V$ should be 0.0065 and 0.009, respectively.
            The ratio of the core radius to the stellar radius, $\sqrt{f_\mathrm{core}}$ is 0.08 in this case.

        \subsection{Fading events before splitting}\label{sec:whole}
          In the previous discussions, we showed that a circumstellar dust clump, a precessing planet, and an accretion hotspot or a dust clump are most likely origins for dip-A, B, and C, respectively.
          Here, we inspect the whole scenario from the single fading events before the split to our observations.

          We try to explain our data of 2014 February 23.
          One mysterious result is the split of the fading event after the date.
          The most intuitive origin of the single fading event is a combination of the dust cloud, which causes dip-A, and the planet, which yields dip-B.
          To confirm this idea, we examine the dust clump hypothesis in sub-subsection~\ref{sec:dust_A}.
          The obtained filling factor of the core is about 0.005, which is consistent with the depth of dip-B.
          Therefore, the precessing planet surrounded by the optically thin dust envelope could have caused the single fading event on 2014 February 23.
          This suggests that the dust component left the planet for some reason (e.g., mass outflow) and caused dip-A later.
          Based on this scenario, it is notable that dip-A survived until 2017 October 3 after dip-B disappeared between 2015 October 17 and 2016 November 25.
          In other words, dust clump, dip-A, does not follow the precessing planet.
          The Hill sphere radius of the planet is calculated as $< 0.158\,a$ with assumptions on the masses of $M_* = 0.39\,M_\odot$ and $M_p < 4.9\,M_\mathrm{Jup}$ (see subsection~\ref{sec:PTFO}).
          It is converted to $< 0.025$ in phase.
          The phase difference between dip-A and B is about 0.1 (figure~\ref{fig:lc_def}), much larger than the Hill sphere radius.
          Therefore, the dust clump is not bound to the planet and it is natural that dip-A shows different behavior from dip-B based on our scenario.

          Next, we examine whether the precessing planet with the dust envelope is consistent with the past observations.
          Because depths of fading events before February 2014 are typically a few percents, attenuation by the dust component would be the dominant source of such fading events.
          Hence, the asymmetric and changing shape of the fading events reported in \citet{VanEyken2012} is not necessarily explained through the gravity-darkening model, as shown by \citet{Barnes2013}, \citet{Kamiaka2015}, and \citet{Howarth2016}.
          Moreover, if the orbit of this dust clump has the same orbit with the planet, its semi-major axis, $a = 3.94\,R_*$, exceeds the dust-sublimation radius $R_s \sim 2.7\,R_*$ reported in \citet{Yu2015}, and therefore, it is likely that the dust clump had survived for three years from 2009 to 2012.

          Last, we show that our scenario is even consistent with the observational results, which were thought as negative evidence to the planetary hypothesis.
          First, if our derived precession period is correct, the non-detections of the RM effect \citep{Ciardi2015,Yu2015} are reasonable because the planet did not transit the star at their observations owing to the small inclination angles (see two vertical lines in figure~\ref{fig:inc}).
          Second, the non-detection of a secondary eclipse reported in \citet{Yu2015} is also reasonable.
          Employing our derived transit parameters in table~\ref{tab:orb_param}, the reflected light and thermal emission from the planet would be 0.05 and 0.26 times weaker than their estimation, respectively.
          This indicates that the expected depth would be shallower than Spitzer's and Magellan's $3\,\sigma$-upper limits of detection (see also figure~8 in \cite{Yu2015}).
          Finally, the fading event obtained by Spitzer \citep{Ciardi2015} can also be explained by this precession scenario.
          Because attenuation by the dust component would be negligible in such infrared data, the fading event detected by Spitzer would reflect only the planetary transit.
          We took Spitzer data from table~1 in \citet{Ciardi2015} and derived the inclination angle by MCMC fitting as the individual fitting described in sub-subsection~\ref{sec:planet}.
          As shown by the red cross in figure~\ref{fig:inc}, our derived sinusoidal function is also consistent with the inclination angle of Spitzer data.

        \subsection{PTFO\,8-8695\,b}\label{sec:PTFO}
          We discuss the characteristics of this putative planet here.
          The planet mass, $M_p \sin i < 4.8\,\pm\,1.2\,M_\mathrm{Jup}$ was obtained by the previous RV measurements by \citet{VanEyken2012}.
          Since the typical inclination angle at their observations is $78^\circ$ based on our precession period and amplitude, an upper limit for the planetary mass comes to $4.9\,\pm\,1.2\,M_\mathrm{Jup}$, which is slightly smaller than $5.5\,M_\mathrm{Jup}$ derived by \citet{VanEyken2012}.
          The planet radius was derived as $0.36\,\pm\,0.15\,R_\mathrm{Jup}$ based on the planet hypothesis.
          The core-halo dust clump hypothesis, on the other hand, allows still smaller radius of about $0.22\,R_\mathrm{Jup}$.
          Our estimates of the planetary radius are much smaller than the previous estimates (e.g., $1.91\,R_\mathrm{Jup}$ by \cite{VanEyken2012}).
          This is because the previous estimates were based on the dips before splitting, whilst our estimates are based on the shallower dip B after splitting.
          This small radius suggest that the putative planet is not a ordinary gas giant planet but is more likely a super-Earth to Neptune-sized planet or a disintegrating former gas giant planet.
          However, only the upper limit information on the planetary mass hinders further discussions.
      
          The two main characteristics of this planet are its youth and relatively small radius.
          From the viewpoint of the youth, this is the third planet candidate younger than a few Myr, as listed in table~\ref{tab:young}.
          \begin{table*}
            \tbl{List of young hot Jupiters and candidates}{
              \begin{tabular}{lccccccc} \hline
                Name & Stellar age & Method & Semi-major axis$\>\mathrm{(au)}$ & Mass$\>(M_\mathrm{Jup})$ & Radius$\>(R_\mathrm{Jup})$ & Reference\footnotemark[$*$] \\ \hline
                PTFO\,8-8695\,b\footnotemark[$\dagger$] & $2.6\>$Myr & Transit & $0.0083\,\pm\,0.0007$ & $< 4.9\,\pm\,1.2$ & $0.36\,\pm\,0.15$ or $\sim0.22$ & (1)(2)(3) \\
                V830\,Tau\,b & $2\>$Myr & RV & $0.057\,\pm\,0.001$ & $0.77\,\pm\,0.15$ & -- & (4)(5) \\
                K2-33\,b & $11\>$Myr & Transit & $0.0409^{+0.0021}_{-0.0023}$ & $<3.6$ & $0.514_{-0.052}^{+0.055}$ & (6)(7) \\
                CI\,Tau\,b\footnotemark[$\dagger$] & $2\>$Myr & RV & $\sim 0.079$ & 11--12 & -- & (8)(9) \\
                TAP\,26\,b & $17\>$Myr & RV & $0.086\,\pm\,0.003$ & $2.03\,\pm\,0.46$ & -- & (10)(11) \\ \hline 
              \end{tabular}}\label{tab:young}
              \begin{tabnote}
                \footnotemark[$*$] (1) \citet{Briceno2005}; (2) \citet{VanEyken2012}; (3)~this work; (4) \citet{Donati2015}; (5)\citet{Donati2016}; (6) \citet{David2016}; (7) \citet{Mann2016}; (8) \citet{Guilloteau2014}; (9) \citet{Johns-Krull2016a}; (10) \citet{Grankin2013}; (11) \citet{Yu2017} \\
                \footnotemark[$\dagger$] Candidates. 
              \end{tabnote}
          \end{table*}
          Because this planet would be too close-in to be formed in situ, it might have been formed in an outer orbit and migrated inward.
          Considering the timescales of the migration mechanisms (i.e., disk-planet interaction, planet-planet scattering, and Kozai--Lindov mechanism), the most likely migration mechanism would be disk-planet interaction.
          The same conclusion as for V830\,Tau\,b \citep{Donati2016} and K2-33\,b \citep{Mann2016} implies that the planet formation and orbital evolution would be generally completed within a few Myr.
          Recently, \citet{Schmidt2016} reported on the dete ction of a putative second planet candidate CVSO\,30\,c (PTFO\,8-8695\,c) orbiting at $662\>\mathrm{au}$.
          If their planet candidate and our identification of a transiting planet are both true, PTFO8-8695 is the first system where a close-in planet and a wide separation planet are found to be orbiting the same star.
          Then, this system is valuable for future studies on planet-planet scattering.
          However, the identification of CVSO\,30\,c is questioned by \citet{Lee2018} because their additional optical photometry is better fit by a background star.
          So, we need to wait for further studies before concluding an origin of migration of the two putative planets.
          
          On the other hand, the relatively small size of this planet would result from the mass outflow concluded by \citet{Johns-Krull2016}, and therefore, the planet at present might be an atmosphere-escaping planet.
          In addition, the split of the fading event into dips-A and B might have been driven by such mass outflow.

  \section{Conclusion}
    We carried out simultaneous optical and infrared observations of PTFO\,8-8695, which may be accompanied by a young transiting planet candidate reported in \citet{VanEyken2012}.
    Through about five-years of follow-up monitoring using the Kanata telescope, we found that there are currently three types of fading events: dip-A at phase $-0.1$, dip-B at phase 0, and dip-C at phase 0.5.
    As a result of examining four hypotheses (i.e., a starspot, an accretion hotspot, a planet, and a dust clump), we conclude that the origins of dips-A and B are a dust cloud and a precessing planet, respectively.
    For dip-C, we can only limit the possible origins to a hotspot and a dust clump.
    Considering the single fading event observed in February 2014, the most likely scenario for this system is that a precessing planet was associated with a dust cloud and then split into dips-A and B.
    This scenario is consistent with previous studies such as the implausibility of the precession model \citep{Howarth2016}, the non-detections of the RM effect \citep{Ciardi2015, Yu2015}, and the Spitzer observation reported in \citet{Ciardi2015}.

    Indeed, continuous observations are needed to make our conclusion more robust, but our observations support the existence of a planet whose mass might be $< 4.9\,\pm\,1.2\,M_\mathrm{Jup}$ and radius might be $0.36\,\pm\,0.15\,R_\mathrm{Jup}$ or about $0.22\,R_\mathrm{Jup}$.
    If this planet really exists, this is the third case of a planet candidate younger than $3\>$Myr and suggests that planets could be formed in such short timescale.
    To confirm the correctness of this precessing model, observations around November 2019, when the orbit would be edge-on, are efficient.
    We believe that accurate follow-up observations would give us a clue to reveal not only the nature of this object but also the planet formation process.

  \begin{ack}
    The authors are grateful to the Kanata telescope team at Hiroshima University for their kind support in performing the observations.
    We also thank Yousuke Utsumi and Yuki Moritani for their contribution to our observations.
    Lastly, we acknowledge the fruitful comments from an anonymous referee, which greatly improved the paper.
  \end{ack}

\end{document}